\documentclass[11pt]{article}
\usepackage{amsmath,amsfonts,amsbsy,amssymb,array,accents,dsfont}
\usepackage{enumerate,array,latexsym,graphicx,mathrsfs,verbatim,psfrag}
\usepackage{bm} %math bold symbols
\usepackage[normalem]{ulem}%strikethrough
\usepackage{multirow}
\usepackage{cellspace}
\usepackage{pdflscape}
\usepackage{booktabs} %some alignment stuff
\usepackage[usenames]{color}
\usepackage[utf8]{inputenc}
\usepackage[english]{babel}
\usepackage{fancybox}

\usepackage{datetime}

\usepackage[nosort]{cite}
\usepackage{chngpage} % allows for temporary adjustment of side margins
\usepackage{setspace}
\usepackage{tensor}

\usepackage{enumitem}
\setlist{itemsep=0pt}

\usepackage[colorlinks=true,      linkcolor=darkblue,      urlcolor=darkblue,      
            filecolor=darkblue,      citecolor=darkblue,       pdfstartview=FitH,     
						pdfpagemode=UseNone,      bookmarksopen=true]{hyperref}  % LINKS
\usepackage[all]{hypcap}     %should be loaded AFTER hyperref, which should otherwise be loaded last.

\newcommand{\captionfonts}{\small}
\makeatletter  % Allow the use of @ in command names
\long\def\@makecaption#1#2{%
  \vskip\abovecaptionskip
  \sbox\@tempboxa{{\captionfonts #1: #2}}%
 \ifdim \wd\@tempboxa >\hsize
    {\captionfonts #1: #2\par}
  \else
    \hbox to\hsize{\hfil\box\@tempboxa\hfil}%
  \fi
  \vskip\belowcaptionskip}
\makeatother   % Cancel the effect of \makeatletter

\DeclareMathSymbol{\medhatsym}{\mathord}{largesymbols}{"62} % basic symbol

\DeclareMathSymbol{\medtildesym}{\mathord}{largesymbols}{"65}% basic symbol

\newcommand{\comm}[1]{} %for commenting out blocks of text

\def\({\left(}
\def\){\right)}
\def\[{\left[}
\def\]{\right]}

\def\One{{\hbox{ 1\kern-.8mm l}}}

\def\barray{\begin{array}}
\def\earray{\end{array}}
\def\be{\begin{equation}}
\def\ee{\end{equation}}
\def\bea{\begin{eqnarray}}
\def\eea{\end{eqnarray}}
\def\bal{\begin{align}}
\def\eal{\end{align}}

\numberwithin{equation}{section} % replaces the hack below

\makeatletter
\g@addto@macro\bfseries{\boldmath}
\makeatother

\definecolor{cardinal}{rgb}{0.6,0,0}
\definecolor{darkgreen}{rgb}{0,0.4,0}
\definecolor{purple}{rgb}{0.5, 0, 0.5}
\definecolor{golden}{rgb}{0.92, 0.7, 0}
\definecolor{midnight}{rgb}{0, 0, 0.5}
\definecolor{darkblue}{rgb}{0, 0, 0.8}

\begin{document}

%\begin{titlepage}

\begin{flushright}
%
%IPHT-T17/134\\
%
\end{flushright}

\vspace{14mm}

\begin{center}

{\huge \bf{D1-D5-P superstrata in 5 and 6 dimensions:\\ separable wave equations and  prepotentials}} \medskip 

%{\huge \bf{Microstate Geometries}}

\vspace{13mm}

%\bigskip\bigskip
\centerline{{\bf  Robert Walker$^{1}$}} 
\bigskip
\bigskip
\vspace{1mm}

\centerline{$^1$\,Department of Physics and Astronomy,}
\centerline{University of Southern California,} \centerline{Los
Angeles, CA 90089-0484, USA}

\vspace{4mm}

{\small\upshape\ttfamily  ~  walkerra @ usc.edu} \\

\vspace{10mm}
%\bigskip\bigskip\bigskip
 
%\textsc{Abstract}

\begin{adjustwidth}{17mm}{17mm} % to adjust the L and R margins
\begin{abstract}
\vspace{3mm}
\noindent

We construct the most general single-mode superstrata in 5 dimensions with ambipolar, two centered Gibbons Hawking bases, via dimensional reduction of superstrata in 6 dimensions. Previously, asymptotically $\text{AdS}_{3}\times \mathbb{S}^{2}$ 5-dimensional superstrata have been produced, giving microstate geometries of black strings in 5 dimensions. Our construction produces asymptotically $\text{AdS}_{2}\times \mathbb{S}^{3}$ geometries as well, the first instances of superstrata describing the microstate geometries of black holes in 5 dimensions. New examples of superstrata with separable massless wave equations in both 5 and 6 dimensions are uncovered. A $\mathbb{Z}_{2}$ symmetry which identifies distinct 6-dimensional superstrata when reduced to 5 dimensions is found. Finally we use the mathematical structure of the underlying hyper-K\"{a}hler bases to produce prepotentials for the superstrata fluxes in 5 dimensions and uplift them to apply in 6 dimensions as well.

\end{abstract}
\end{adjustwidth}

\end{center}

%\end{titlepage}

\thispagestyle{empty}

\newpage

%%%%%%%%%%%%%%%%%%%%%%%%%%%%%%%%%%%%%

\baselineskip=14pt
\parskip=2pt

\tableofcontents

% \newpage

\baselineskip=15pt
\parskip=3pt

%%%%%%%%%%%%%%%%%%%%%%%%%%%%%%%%%%%%%
\section{Introduction}
\label{Sect:Intro}
%%%%%%%%%%%%%%%%%%%%%%%%%%%%%%%%%%%%%
The microstate geometry program seeks to describe black hole entropy by explicitly constructing smooth horizonless geometries that approximate a given black hole \cite{Bena:2007kg}. These geometries are interpreted as microstates in the ensemble of states that give rise to the entropy via a Boltzmann like state counting procedure\footnote{The foundational work of \cite{Strominger:1996sh} performed a state counting of this form in a regime where the black hole geometry is absent. The microstate geometry program seeks to describe such states in the regime where the gravitational geometry is manifest.}. This idea has been explored most fully in the D1-D5-P system of type IIB supergravity. There are two main research directions in this program, to produce new examples of micrsostate geometries and to better understand those that we already have. This work was motivated by finding new examples of superstrata with separable massless wave equations (SMWEs), a property that has proven to be critical in recent analysis and critiques of the microstate geometry program  \cite{Bena:2017upb,Raju:2018xue,Tyukov:2017uig,Heidmann:2019zws,Bena:2018bbd,Bena:2019azk} .

We focus on the microstate geometries that have come to be known as superstrata \cite{Heidmann:2019zws,Bena:2011uw,Bena:2015bea,Ceplak:2018pws}. These solutions have several key features that make them ideal for exploring the microstate geometry program:
\begin{itemize}
\item The geometry can be produced with either asymptotically flat, or asymptotically anti-de Sitter crossed with a sphere \cite{Bena:2017xbt} (possibly with orbifold singularities). 
\item They can be tuned to produce arbitrarily long BTZ-like throats prior to smoothly capping off \cite{Tyukov:2017uig}.
\item It is known how to construct them in both 5 and 6 dimensions \cite{Bena:2017geu}. 
\item There are families of solutions with the same asymptotic charges \cite{Heidmann:2019zws}. 
\item Examples can be produced with greater coverage of the charges \cite{Bena:2016ypk}. For instance, earlier constructions such as in \cite{Bena:2005va,Berglund:2005vb,Bena:2006kb,Bena:2007qc} could only produce high angular momentum solutions, there is no such obstruction for superstrata. 
\item Some examples are known to have SMWEs \cite{Bena:2017upb}, this allows the computation of properties such as energy gaps \cite{Tyukov:2017uig} in the spectrum or investigation of scattering \cite{Heidmann:2019zws,Bena:2018bbd,Bena:2019azk}. 
\item The dual CFT description \cite{Bena:2016agb} is well understood.
\end{itemize}
It is for these reasons that the superstrata have risen to prominence, with many recent investigations \cite{Bena:2019azk,Giusto:2019qig,Tian:2019ash,Bena:2018mpb,Bakhshaei:2018vux,Tormo:2019yus}. 

The \textit{original} superstrata constructed in \cite{Bena:2015bea} have three important generalizations that need to be distinguished. To begin with the original superstrata were generated by solely bosonic CFT operators, the work of \cite{Ceplak:2018pws} introduced fermionic operators to produce \textit{supercharged} superstrata. In \cite{Heidmann:2019zws} a superposition of the original and supercharged superstrata gave \textit{hybrid} superstrata, steps were also taken towards constructing superpositions of solutions with multiple modes. Throughout this work we will refer to all of these solutions as superstata, distinguish between the separate flavors (original, supercharged, hybrid) when required and treat single and multi-mode solutions separately.   

The defining feature of superstrata is they allow fluctuations in the Maxwell fields along the periodic coordinates. In 6 dimensions the fluctuations are parametrized by three integers $(k,m,n)$ corresponding to Fourier modes for the three periodic coordinates $(v,\phi,\psi)$. The 6-dimensional superstrata can be expressed as a double circle fibration in the coordinates $(v,\psi)$. A natural $\text{SL}(2,\mathbb{Q})$ action, known as a spectral transformation \cite{Bena:2008wt,Niehoff:2013kia}, can be defined which mixes these circles. Any single-mode 6-dimensional superstrata will be cyclic in some combination of the $(v,\psi)$ circles, so a Kaluza-Klein reduction on this combination of circles is possible. In order to preserve the form of the BPS equations in 5 dimensions it is useful to use a spectral transformation redefining the $(v,\psi)$ coordinates so that the cyclic direction becomes exactly $v$. 

In addition to the integers $(k,m,n)$, the $\text{SL}(2,\mathbb{Q})$ transformation introduces another 3 parameters, giving a total of 6 parameters. One of these parameters is used to ensure the reduction occurs on the $v$-circle. The remaining 5 parameters then show up in the 5-dimensional solutions as: 2 Fourier modes for the $(\phi,\psi)$ directions, the 2 Gibbons Hawking (GH) charges of the now two centered ambipolar GH base and 1 gauge degree of freedom. Thus the reduction produces the most general single-mode superstrata possible on an ambipolar two centered GH base in 5 dimensions. If the net GH charge vanishes the asymptotic geometry is $\text{AdS}_{3}\times \mathbb{S}^{2}$, such geometries were produced in \cite{Bena:2017geu} and correspond to microstate geometries for black strings. If the net GH charge is non-zero the  asymptotic geometry is $\text{AdS}_{2}\times \mathbb{S}^{3}$ with a possible $\mathbb{Z}_{p}$ orbifolding of the $\mathbb{S}^{3}$, these are the microstate geometries of black holes, a new result.

We use spectral transformations and reductions to produce new examples of superstrata with SMWEs. In addition to the original $(1,0,n)$ family that were known to have SMWEs \cite{Bena:2017upb}, we show that the $(1,1,n)$ family do as well in 6 dimensions. Applying spectral transformations to these families we find that the two remaining spectral transformation parameters index families of distinct 6-dimensional superstrata with SMWEs. The parameters can be used to alter the complexity of the individual separated differential equations. In addition we show that the $(2,1,n)$ family has SMWEs in certain circumstances: in 6-dimensions the supercharged flavor have SMWEs \cite{Heidmann:2019zws}, in 5-dimensions both the supercharged and original flavors have SMWEs, while the hybrid flavor have SMWEs in 5-dimensions provided the momentum on the $\phi$-circle vanishes.

We show that the $(k,m,n)$ and $(k,k-m,n)$ superstrata in 6 dimensions reduce to the same solutions in 5 dimensions, hence there is a $\mathbb{Z}_{2}$ symmetry identifying 6-dimensional solutions after reduction. In addition it is also clear that multi-mode solutions will not reduce unless the multiple modes are parallel in the $(v,\psi)$ directions. Hence we reveal two mechanisms that may lead to a greater number of superstrata in 6 than 5 dimensions.

In \cite{Tyukov:2018ypq} it was shown how how in 5 dimensions the superstrata fluxes could be derived from a scalar \textit{prepotential}. This prepotential program is of interest since it promises to simplify the process of finding BPS solutions to the D1-D5-P system by reducing parts (if not all) of it to functional analysis on 4-dimensional hyper-K\"{a}hler bases. We construct the prepotentials for our new 5-dimensional superstrata, as well as indicate how the reduction procedure can be inverted so that prepotentials can be used in 6 dimensions as well.

In section \ref{Sect:Sec2 superstrata intro} we give an overview of the superstata solutions, including the BPS equations they solve and how they are constructed. The following sections then separate four related sets of original results:
\begin{itemize}
\item Section \ref{Sect:Sec3 relating 5D 6D superstrata} illustrates the relationship between single-mode superstrata in 6 and 5 dimensions using spectral transformations. 
\item Section \ref{Sect:Sec4 relations amongst superstrata} shows that dimensional reduction of the $(k,m,n)$ and $(k,k-m,n)$ 6-dimensional superstrata leads to equivalent 5-dimensional superstata. The special case of the $(1,0,n)$ and $(1,1,n)$ families is considered explicitly.
\item  Section \ref{Sect:Sect5 Seperability} summarizes a non-exhaustive but systematic search for superstrata with SMWEs, we show how the $(2,1,n)$ families has greater separability properties in 5 than 6 dimensions and how spectral transformations can alter the form of the wave equations in 6 dimensions. 
\item Section \ref{Sect 3.5: prepotentials} shows how prepotentials can be constructed for superstrata fluxes in both 5 and 6 dimensions, explicit examples are given.  
\end{itemize}
Finally, a discussion of the significance of these results and possible directions for future investigation is given in section \ref{Sect:Sect6 Discussion}.

%%%%%%%%%%%%%%%%%%%%%%%%%%%%%%%%%%%%%
\section{Superstrata and their flavors in supergravity} 
\label{Sect:Sec2 superstrata intro}
%%%%%%%%%%%%%%%%%%%%%%%%%%%%%%%%%%%%%
This section reviews the BPS equations in 6 dimensions and sketches how to construct the superstrata, more details may be found in \cite{Heidmann:2019zws,Bena:2015bea,Ceplak:2018pws,Bena:2017xbt}.

\subsection{BPS equations}
The superstrata and its flavors constructed in \cite{Heidmann:2019zws,Bena:2015bea,Ceplak:2018pws} are generally studied within 6-dimensional (0,1) supergravity obtained by compactifying type IIB supergravity with manifold structure $\mathcal{M}^{1,4}\times \mathbb{S}^{1}\times \mathcal{C}$ on $\mathcal{C}$. The compactification manifold $\mathcal{C}$ is required to be hyper-K\"{a}hler, thus it is taken to be either $\mathbb{T}^{4}$ or K3. The circle $ \mathbb{S}^{1}$ of radius $R$ is paramatrized by the cyclic coordinate
\begin{align}
y \sim y+2 \pi R~.
\end{align}
The simplest models that give smooth superstrata involve coupling to two tensor multiplets. It is also possible \cite{Bena:2017geu}, in certain circumstances to compactify the theory on a circle direction inside the $\mathcal{M}^{1,4}\times \mathbb{S}^{1}$, the theory then reduces to a 5-dimensional $\mathcal{N}=2$ supergravity coupled to three vector multiplets. This compactification is nothing more than a standard Kaluza-Klein reduction, which ensures the BPS equations in each dimension are related.

The 6-dimensional geometry can be written as
\begin{align}
ds_6^2 &=   -\frac{2}{\sqrt{P}} \, (dv+\beta) \big(du +  \omega + \tfrac{1}{2}\, F \, (dv+\beta)\big) 
+  \sqrt{P} \, ds_4^2(\mathcal{B})~, \label{ds6} \\
&= \frac{1}{F\sqrt{P}}\left( (du+\omega)^{2}+FPV \,ds_{3}^{2}\right) -\frac{F}{\sqrt{P}}\left( dv+ \beta +\frac{1}{F}(du+\omega)\right)^{2} +\frac{\sqrt{P}}{V}\left(d\psi +A \right)^{2}~.\label{dsDoubleFiber}
\end{align}
where\footnote{Often the (perhaps) more canonical pair of light cone coordinates $u=\frac{1}{\sqrt{2}}(t-y)$ and $v= \frac{1}{\sqrt{2}}(t+y)$ are used instead of (\ref{uvDef}). It is shown in \cite{Bena:2017geu} how the two choices are related by a redefinition of the $(F,\omega,\Theta^{(I)})$. Here we use the $t=u$ identification since since when we compactify on $v$ to produce 5-dimensional solutions, $u=t$ will indeed be the time direction.} 
\begin{align}
u= t ~, \qquad v=t+y~, \label{uvDef}
\end{align}
and the details of $(ds_{4}^{2}(\mathcal{B}),ds_{3}^{2})$ are discussed around (\ref{GHbase}). Supersymmetry requires all fields, such as the functions $(P,F)$, one form $\beta$ and the $(Z_{I},\Theta^{(I)})$ to be independent of $u$. Working with $v$ independent $\beta$ and $ds_4^2(\mathcal{B})$ simplifies the BPS equations as well, demanding this ensures $ds_4^2(\mathcal{B})$ is hyper-K\"{a}hler and $d \beta$ is self dual on this base.

The form of the metric in (\ref{dsDoubleFiber}) is that of a double circle fibration in for the $(v,\psi)$ circles, thus there is a natural $\text{SL}(2,\mathbb{Z})$ action redefining the $(v,\psi)$ coordinates amongst each other. This action may be used to ensure the fields and metric are independent of $v$, the 5-dimensional solution is then found by applying a Kaluza Klein reductions to the $v$-circle. Completing this procedure and identifying 
\begin{align}
F=-Z_{3} ~,\label{6to5data}
\end{align}
gives the 5-dimensional geometry 
\begin{align}
ds_{5}^{2} &= \left(Z_{3}P \right)^{-\frac{2}{3}} (dt+\omega)^{2} + \left( Z_{3}P\right)^{\frac{1}{3}} \, ds_4^2(\mathcal{B}) ~.\label{ds5}
\end{align} 

The full supergravity system has a set of maxwell like fields $(Z_{I},\Theta^{(I)})$ where $I\in \{1,2,4 \}$, in term of which the field strengths in of the vector/tensor multiplets can be written (See \cite{Bena:2017geu} for instance). These Maxwell like fields and the data appearing in (\ref{ds6}) and (\ref{ds5}) are fixed by the BPS equations. In 6-dimensions the BPS equations split into a first layer:
\begin{align}
 * D  \dot{Z}_1 &=  D \Theta^{(2)} \,,\quad D* D Z_1 = -\Theta^{(2)} \wedge d\beta\,,\quad \Theta^{(2)} =* \Theta^{(2)} ~, \label{eqZ1Theta2} \\
* D \dot{Z}_2  &=  D \Theta^{(1)} \,,\quad  D * D Z_2 = -\Theta^{(1)} \wedge d\beta\,,\quad \Theta^{(1)} =* \Theta^{(1)}~,\label{eqZ2Theta1} \\
* D \dot{Z}_4 &=  D  \Theta^{(4)} \,,\quad D * D Z_4 = - \Theta^{(4)}\wedge d\beta\,,\quad  \Theta^{(4)}=*  \Theta^{(4)} ~,\label{eqZ4Theta4}
\end{align}
as well as a second layer:
\begin{align}
(1+*)D\omega +F\,d\beta &= Z_{1}\Theta^{(1)} +Z_{2}\Theta^{(2)}-2Z_{4}\Theta^{(4)}~, \label{BPS6D1}\\
*D* \left( \dot{\omega}- \frac{1}{2}DF \right) &= \ddot{P}- \left(\dot{Z}_{1}\dot{Z}_{2}-\dot{Z}^{2}_{4} \right) - \frac{1}{2}* \left(\Theta^{(1)} \wedge \Theta^{(2)} - \Theta^{(4)} \wedge \Theta^{(4)} \right)~,\label{BPS6D2}
\end{align}
where $(d,*)$ are the exterior derivative and Hodge star operations on $ds_4^2(\mathcal{B})$, a dot denotes differentiation with respect to $v$ and 
\begin{align}
D\Phi = d\Phi - \beta \wedge \dot{\Phi}~.
\end{align}
The function $P$ is fixed by 
\begin{align}
P = Z_{1}Z_{2}-Z_{4}^{2}~.
\end{align}

If the data $(Z_{I},\Theta^{(I)},\beta,F,\omega)$ appearing above are independent of $v$ then the 6-dimensional BPS equations after defining 
\begin{align}
d\beta = \Theta^{(3)}~,
\end{align}
reduce to the 5-dimensional BPS equations, with zeroth layer:
\begin{align}
\Theta^{(I)} =* \Theta^{(I)} ~, \qquad \Theta^{(3)} = *\Theta^{(3)}~, \label{5D Theta dual}
\end{align}
first layer:
\begin{align}
\nabla^{2} Z_{1} &= * \left(\Theta^{(2)}\wedge \Theta^{(3)} \right)~,\label{5D BPS Z1eq}\\
\nabla^{2} Z_{2} &= * \left(\Theta^{(1)}\wedge \Theta^{(3)} \right) ~,\\
\nabla^{2} Z_{3} &= * \left(\Theta^{(1)}\wedge \Theta^{(2)}-\Theta^{(4)}\wedge \Theta^{(4)} \right)~,\\
\nabla^{2} Z_{4} &= * \left(\Theta^{(3)}\wedge \Theta^{(4)} \right)~,\label{5D BPS Z4eq}
\end{align}
and second layer
\begin{align}
(1+*)dw = Z_{1}\Theta^{(1)}+Z_{2}\Theta^{(2)}+Z_{3}\Theta^{(3)} -2 Z_{4}\Theta^{(4)}~. \label{5DBPSfinal}
\end{align}
It is key to note that in order for this reduction to work all 6-dimensional fields including the $(Z_{I},\Theta^{(I)})$ must be independent of the  $v$-circle that we reduce on. This will be critical in section \ref{SubSec: 6D5D relationship} where we illustrate the relationship between 6 and 5-dimensional superstrata. This is also the reason we need to introduce spectral transformations in section \ref{SubSec: spectral flow}, which will enable a transformation of any given single-mode 6-dimensional superstrata to remove all $v$-dependence before reducing to 5 dimensions. 

\subsection{Gibbons Hawking bases}
The first step in finding solutions to the BPS equations (\ref{eqZ1Theta2})-(\ref{BPS6D2}) or (\ref{5D Theta dual})-(\ref{5DBPSfinal}) is to specify a hyper-K\"{a}hler base. The Gibbons Hawking (GH) geometries provide some of the simplest yet non-trivial examples of hyper-K\"{a}hler manifolds. They are constructed as 
\begin{align}
ds_4^2(\mathcal{B}) = \frac{1}{V}\left( d\psi +A \right)^{2} +V \, ds_{3}^{2}~, \qquad \nabla_{3}^{2}V=0 ~, \qquad *_{3}d_{3}V=d_{3}A~, \label{GHbase}
\end{align}
where $\psi \in [0,4\pi)$ is the GH fiber, $ ds_{3}^{2}$ is the flat metric of $\mathbb{R}^{3}$ and operations with a subscript $_3$ refer to this flat base\footnote{In this paper we use $(\nabla^{2},*,d)$ to refer to the Laplace-Beltrami operator, Hodge star and exterior derivative on the entire 4-dimensional GH base of (\ref{GHbase}).}. Introducing Cartesian coordinates $(y^{1},y^{2},y^{3})$ on $ ds_{3}^{2}$, the $V$ appearing in (\ref{GHbase}) are then given by
\begin{align}
V(\vec{y}) = \sum_{i=1}^{N} \frac{q_{i}}{\left| \vec{y}-\vec{y}_{i} \right|}~,
\end{align}
where the $q_{i}\in \mathbb{Z}$ are known as the GH charges, they are centered at the $\vec{y}_{i}$ and $N$ labels the total number of charges. 

For computations it is convenient to introduce spherical bipolar coordinates $(r,\theta,\phi)$ on the flat $\mathbb{R}^{3}$ defined by 
\begin{align}
y_{1}+iy_{2} = \frac{r}{4}\sqrt{r^{2}+a^{2}}\sin 2\theta \, e^{i\phi} \qquad \text{and} \qquad y_{3} = \frac{1}{8}(2r^{2}+a^{2})\cos 2\theta~,
\end{align}
where $r\in [0,\infty$, $\theta \in [0,\pi/2)$ and $\phi \in [0,2\pi)$. Defining 
\begin{align}
V &= \frac{4}{\Lambda} ~, \qquad\qquad\qquad~~\, A = \frac{(a^{2}+2r^{2})\cos 2\theta -a^{2}}{2\Lambda } \, d\phi~, \\
\Sigma &= r^{2}+a^{2}\cos^{2}\theta~, \qquad \Lambda= r^{2}+a^{2}\sin^{2}\theta~, \label{SigmaLambda}
\end{align}
the GH metric then becomes
\begin{align}
ds_{4}^{2}(\mathcal{B}) = \frac{1}{V}(d\psi+A)^{2} + \frac{V}{16} \left( 4\Sigma \Lambda \left( \frac{dr^{2}}{a^{2}+r^{2}}+d\theta^{2} \right) + r^{2}(r^{2}+a^{2})\sin^{2}2\theta \, d\phi^{2} \right)~. \label{GHsbcoords}
\end{align}
These coordinates are adapted to the superstrata since the $(Z_{I},\Theta^{(I)})$ are sourced on the locus $\Sigma=0$. 

In model building it is important to understand that the $\Theta^{(I)}$ are supported on 2 cycles in the geometry. In the standard construction of 6-dimensional superstrata the base is taken to be flat $\mathbb{R}^{4}$. The non-trivial 2 cycles are then provided by the pinching off of the $v$-circle. In 5 dimensions the non-trivial 2 cycles are contained entirely in the GH base, they are furnished by the pinching off of the $\psi$-circle where $V$ diverges at the GH points. Thus the 5-dimensional superstrata require GH bases with multiple centers. The simplest such geometries have two centers, which can be aligned with the $y^{3}$ direction and their separation parametrized by $a$
\begin{align}
\vec{y}_{\pm}=(0,0,\pm a^{2}/8)~.
\end{align}
This gives
\begin{align}
V=\frac{q_{-}}{r_{-}}+\frac{q_{+}}{r_{+}}~,
\end{align}
where
\begin{align}
 r_{-}=\left| \vec{y}- \vec{y}_{-}\right|=\frac{\Sigma}{4}\qquad \text{and} \qquad r_{+}=\left| \vec{y}- \vec{y}_{+}\right|=\frac{\Lambda}{4}~,
\end{align}
and $(q_{-},q_{+})$ are the GH charges.

Naively the $\Theta^{(I)}$ should be the cohomological duals to the second homology on the GH base. Thus it would seem rather pathological to allow $q_{i}$ with varying signs, since the two cycles would be destroyed by the behavior at zeros\footnote{The full 5-dimensional geometry in (\ref{ds5}) is regular at these points, due to the behavior of $Z_{3}P$.} of $V$. However, the key to constructing the superstrata is to allow exactly these types of bases which have come to be known as \textit{ambipolar}\footnote{Ambipolar geometries can be characterized as those geometries that possess domains where the signature is $(+,+,+,+)$ and domains where it is $(-,-,-,-)$, the surfaces where it flips are known as ambipolar surfaces.}. The $\Theta^{(I)}$ fluxes that are then used have a far richer structure than those that can be built out of just the fluxes dual to the two cycles of non-ambipolar bases. Currently a complete understanding of the fluxes that can be constructed on ambipolar bases is missing, it is hoped the prepotential results of section \ref{Sect 3.5: prepotentials} will be helpful for future investigations in this direction, constructing new fluxes and microstate geometries from them. 

For future reference we note that frames on (\ref{GHsbcoords}) can be erected as
\begin{align}
e_{1}=\sqrt{\frac{\Sigma}{r^{2}+a^{2}}}\, dr , ~~ e_{2}=\sqrt{\Sigma}\, d\theta, ~~ e_{3}=\frac{1}{2}\sqrt{a^{2}+r^{2}}\sin\theta \, \left(d\psi-d\phi \right), ~~ e_{4}= \frac{1}{2} r\cos\theta \, \left(d\psi+d\phi \right)~,
\end{align} 
in terms of which a basis for self dual forms is 
\begin{align}
\Omega^{(1)}&= \frac{1}{\Sigma \sqrt{r^{2}+a^{2}}\cos \theta}\left( e_{1}\wedge e_{2}+e_{3}\wedge e_{4}\right)~,\label{Omega1} \\
\Omega^{(2)}&= \frac{1}{\sqrt{\Sigma} \sqrt{r^{2}+a^{2}}\cos \theta}\left( e_{1}\wedge e_{4}+e_{2}\wedge e_{3}\right)~,\label{Omega2} \\
\Omega^{(3)}&= \frac{1}{\sqrt{\Sigma} r\sin \theta}\left( e_{1}\wedge e_{3}-e_{2}\wedge e_{4}\right)~,\label{Omega3}
\end{align}
and the canonical complex structure is 
\begin{align}
J = \frac{1}{\sqrt{\Sigma} r\sin \theta}\left( e_{1}\wedge e_{3}-e_{2}\wedge e_{4}\right) ~. \label{J3}
\end{align}

\subsection{Original, supercharged and hybrid superstrata in 6-dimensions}
Since the hybrid superstrata encompass both the original and the supercharged flavors they provide a convenient way to study the properties of both at once. Hence we will work with the hybrid flavor and fix the relevant parameters to highlight the original or supercharged results where necessary. 

To construct/summarize the $(Z_{I},\Theta^{(I)})$ that solve the 6-dimensional BPS equations (\ref{eqZ1Theta2})-(\ref{BPS6D2}) on the flat $\mathbb{R}^{4}$ base it is convenient to introduce the mode functions
\begin{align}
v_{k,m,n} & \equiv (m+n)\frac{v}{R}-\frac{k}{2}\phi+\frac{1}{2}(k-2 m)\psi~,  \label{moding} \\
\Delta_{k,m,n} & \equiv \left( \frac{a}{\sqrt{r^{2}+a^{2}}}\right)^{k} \left( \frac{r}{\sqrt{r^{2}+a^{2}}}\right)^{n}\cos^{m}\theta \sin^{k-m}\theta~, \label{DeltaDef}
\end{align}
where $(k,m,n)$ are non negative integers indexing Fourier modes on $(v,\phi,\psi)$. The $(Z_{I},\Theta^{(I)})$ will depend non-trivially on these modes. However, a key feature of superstrata is that the mode dependence cancels out in the metrics (\ref{ds6}) and (\ref{ds5}) due to a process known as \textit{coiffuring} (\ref{coiff}). It is also convenient to introduce the functions and forms
\begin{align}
z_{k,m,n} &\equiv \sqrt{2}R \frac{\Delta_{k,m,n}}{\Sigma} \cos v_{k,m,n} ~,\\ 
\vartheta_{k,m,n} & \equiv -\sqrt{2} \Delta_{k,m,n} \left[ \left( (m+n)r\sin\theta+n\left( \frac{m}{k}-1 \right) \frac{\Sigma}{r\sin \theta} \right) \Omega^{(1)}\sin v_{k,m,n} \right. \label{thetatilde} \\
&\qquad\qquad\qquad\qquad\qquad +\left. \left( m \left(\frac{n}{k}+1 \right)\Omega^{(2)} + \left( \frac{m}{k}-1\right)n \Omega^{(3)}\right)\cos v_{k,m,n}\right]~,\notag \\
\varphi_{k,m,n} & \equiv \sqrt{2}\Delta_{k,m,n} \left[\frac{\Sigma}{r\sin\theta} \Omega^{(1)} \sin v_{k,m,n}  + \left(\Omega^{(2)}+\Omega^{(3)}\cos v_{k,m,n} \right)\right]~. \label{thetahat}
\end{align}
The supertrata $Z_{I}$ can then be succinctly summarized as 
\begin{align}
Z_{1} &= \frac{Q_{1}}{\Sigma} + b_{1} \frac{R}{\sqrt{2} Q_{5}}z_{2k,2m,2n}~,\label{Z1} \\
Z_{2}&= \frac{Q_{5}}{\Sigma}~, \label{Z2}\\
Z_{4} &= b_{4} z_{k,m,n}~, \label{Z4}
\end{align}
and the $\Theta^{(I)}$ as 
\begin{align}
\Theta^{(1)} &= 0 ~, \label{Theta1}\\
\Theta^{(2)} &= \frac{R}{\sqrt{2} Q_{5}} \left( b_{1} \vartheta_{2k,2m,2n} + c_{2} \varphi_{2k,2m,2n} \right) ~, \label{Theta2}\\
\Theta^{(4)} &= b_{4}\vartheta_{k,m,n} +c_{4} \varphi_{k,m,n} ~, \label{Theta4}
\end{align}
where $(b_{1},b_{4},c_{2},c_{4})$ are constants. 

It is straightforward to check that equations (\ref{Z1})-(\ref{Theta4}) solve the BPS first layer equations (\ref{eqZ1Theta2})-(\ref{eqZ4Theta4}) with
\begin{align}
\beta = -\frac{R a^{2}}{2 \Sigma}\left(d\phi +\cos 2\theta \, d\psi \right)~.
\end{align}
Difficulties arise when trying to solve the BPS second layer equations (\ref{BPS6D1}) and (\ref{BPS6D2}), due to the quadratic sources. A process known as coiffuring was developed to deal with these difficulties \cite{Bena:2014rea}, ensuring regularity of the gauge fields and geometry. Coiffuring requires
\begin{align}
b_{1}=b_{4}^{2} \qquad \text{and} \qquad c_{2}=2 b_{4}c_{4}. \label{coiff} 
\end{align}
The solution for $(\omega,F)$ in equations (\ref{BPS6D1}) and (\ref{BPS6D2}) is now involved, but algorithmic, we now summarize the solution method (full details can be found in \cite{Bena:2017xbt}). 

One first breaks $(\omega,F)$ into pieces that depend on the mode (\ref{moding}) and those that don't
\begin{align}
\omega &= \omega_{0} + \mu_{k,m,n}\,d\psi+\zeta_{k,m,n}\, d\phi \\
F &= 0 + F_{k,m,n}
\end{align}
where $(\mu_{k,m,n},\zeta_{k,m,n},F_{k,m,n})$ are only functions of $(r,\theta)$ by virtue of the coiffuring (\ref{coiff}) and 
\begin{align}
\omega_{0}=\frac{R a^{2}}{2\Sigma} \left( \cos 2\theta ~ d\phi +d\psi\right)~. \label{omega0}
\end{align} 
Substituting these expressions into the BPS equations (\ref{BPS6D1}) and (\ref{BPS6D2}) gives three independent equations. In principal there should be four, three coming from decomposing (\ref{BPS6D1}) into its self dual pieces and one from (\ref{BPS6D2}), but one of the self dual pieces turns out to vanish identically. Taking combinations of these three equations shows that $\mu_{k,m,n}$ and $F_{k,m,n}$ both satisfy Laplace type equations with non-trivial sources, in each case the problem can be reduced to summing solutions of the DE
\begin{align}
\nabla^{2} \mathcal{F}_{2k,2m,2n} = \frac{\Delta_{2k,2m,2n}}{(r^{2}+a^{2})\Sigma \cos^{2}\theta } ~.
\end{align}
The solution of this DE is given by
\begin{align*}
\mathcal{F}_{2k,2m,2n}= - \sum_{j_{1},j_{2},j_{3}=0}^{j_{1}+j_{2}+j_{3}\leq k+n-1} {
j_{1}+j_{2}+j_{3} \choose j_{1},j_{2},j_{3} } \frac{{
k+n-j_{1}-j_{2}-j_{3}-1 \choose k-m-j_{1},m-j_{2}-1,n-j_{3} }^{2}}{{
k+n-1 \choose k-m,m-1,n
}^{2}} \frac{\Delta_{2(k-j_{1}-j_{2}-1),2(m-j_{2}-1),2(n-j_{3})}}{4(k+n)^{2}(r^{2}+a^{2})}
\end{align*}
where 
\begin{align}
{
j_{1}+j_{2}+j_{3} \choose j_{1},j_{2},j_{3}
} = \frac{(j_{1}+j_{2}+j_{3})!}{j_{1}!j_{2}!j_{3}!}~.
\end{align}
The solutions for $(F_{k,m,n},\mu_{k,m,n})$ can be summarized as
\begin{align}
F_{k,m,n}&= 4 \left[ \left(\frac{m(k+n)}{k}b_{4}-c_{4} \right)^{2} \mathcal{F}_{2k,2m,2n}+ \left(\frac{n(k-m)}{k}b_{4}+c_{4} \right)^{2}\mathcal{F}_{2k,2m+2,2n-2}\right]  ~,  \\
\mu_{k,m,n}&= R \left[ \left( \frac{(k-m)(k+n)}{k}b_{4}+c_{4}\right)^{2}\mathcal{F}_{2k,2m+2,2n} + \left(\frac{mn}{k}b_{4}-c_{4} \right)^{2} \mathcal{F}_{2k,2m,2n-2} -\frac{b_{b^{2}\Delta_{2k,2m,2n}}}{4\Sigma} \right] \notag\\
& \qquad\qquad\qquad\qquad\qquad\qquad\qquad\qquad\qquad\qquad -R\frac{r^{2}+a^{2}\sin^{2}\theta}{4\Sigma} F_{k,m,n} + \frac{R B^{2}}{4\Sigma}    ~, \label{mu eq}
\end{align}
where the term proportional to the constant $B^{2}$ is an arbitrary (for now) homogeneous term. 

Solving for the $\zeta_{k,m,n}$ requires integrating two first order DEs of the form
\begin{align}
\partial_{r} \zeta_{k,m,n} &=  S_{r}\left(r,\theta,F_{k,m,n},\partial_{r}\mu_{k,m,n},\partial_{\theta}\mu_{k,m,n} \right)    ~,\\
\partial_{\theta}  \zeta_{k,m,n} &= S_{\theta}\left(r,\theta,F_{k,m,n},\partial_{r}\mu_{k,m,n},\partial_{\theta}\mu_{k,m,n} \right)  ~,
\end{align}
where $ S_{r}$ and $ S_{\theta}$ are functionals of the given arguments only. Unfortunately the full solution is not known in closed form, but for a given $(k,m,n)$ it is straightforward to perform the integration. For specific sub-families it is possible to solve these equations in closed form. For instance this has been done for the original flavor in the $(1,0,n)$ and $(2,1,n)$ families \cite{Bena:2017upb} and the $(k,0,1)$ family \cite{Bena:2018mpb}. In section \ref{SubSec: 11n} we will present the $(\mu_{1,1,n},\zeta_{1,1,n},F_{1,1,n})$ of the $(1,1,n)$ original superstrata in closed form as well. 

The constant $B$ introduced in (\ref{mu eq}) is used to ensure regularity at $(r=0,\theta=0)$ by fixing $\mu_{k,m,n}(r=0,\theta=0)=0$, this is done by setting
\begin{align}
B^{2} = \frac{b_{4}^{2}+ \frac{k^{2}}{mn(k-m)(k+n)}c_{4}^{2} }{{k \choose m} {k+n-1 \choose n}} = b^{2}+c^{2}
\end{align}
where 
\begin{align}
b= \frac{b_{4}}{\sqrt{{k \choose m} {k+n-1 \choose n}}} \qquad \text{and} \qquad c= \frac{kc_{4}}{\sqrt{mn(k-m)(k+n){k \choose m} {k+n-1 \choose n}}}~.
\end{align}
Demanding regularity at $(r=0,\theta=\pi/2)$ also fixes 
\begin{align}
\frac{Q_{1}Q_{5}}{2R^{2}} = a^{2}+ \frac{b^{2}+c^{2}}{2}~.
\end{align}
The beauty of the hybrid solutions is now evident when one considers the conserved charges 
\begin{align}
J_{R}&= \frac{R}{\sqrt{2}} \left(a^{2}+\frac{m}{k}(b^{2}+c^{2}) \right)~, \qquad J_{L}=\frac{R}{\sqrt{2}}a^{2}~, \qquad Q_{P}= \frac{m+n}{2k}(b^{2}+c^{2})~, \qquad Q_{1,5}~.
\end{align}  
If one were restricted to just the original flavor ($c=0$) or the supercharged flavor ($b=0$), then superstrata with different $(k,m,n)$ and $b$ or $c$ would posses different asymptotic charges. However, with the hybrid flavor we can define 
\begin{align}
b\equiv B \cos \alpha ~, \qquad c\equiv B \sin \alpha~,
\end{align} 
where $\alpha \in [0,2\pi)$ and $B>0$. The parameter $\alpha$ then parametrizes a continuous family of superstrata solutions with identical asymptotic charges.

Finally, we need to consider the restrictions on the integers $(k,m,n)$. For the original superstrata the constraints are $1\leq k$, $0\leq m \leq k$ and $1\leq n$. While for the supercharged and hybrid solutions the requirements are $1\leq m \leq k-1$ and $1\leq n$. Thus when we consider the $(1,0,n)$ and $(1,1,n)$ families in sections \ref{Sect:Sec4 relations amongst superstrata} and \ref{Sect:Sect5 Seperability} we are necessarily looking at the original flavor, but when we look at the $(2,1,n)$ family we consider all three flavors.

%%%%%%%%%%%%%%%%%%%%%%%%%%%%%%%%%%%%%
\section{Relating superstrata in 5 and 6 dimensions} 
\label{Sect:Sec3 relating 5D 6D superstrata}
%%%%%%%%%%%%%%%%%%%%%%%%%%%%%%%%%%%%%
This section shows how spectral transformation can be used to turn any given 6-dimensional single-mode superstrata into a form in which it is independent of $v$. The transformation corresponds to a coordinate redefinition among $(v,\psi)$, followed by a lattice re-identification. It has has a non-trivial effect on the $ds_{4}(\mathcal{B})$ base, turning a flat $\mathbb{R}^{4}$ into an ambipolar two centered GH space. We also discuss how this reduction procedure fails for multi-mode solutions with non-parallel modes.

\subsection{Summary of spectral transformations} \label{SubSec: spectral flow}
Spectral transformations as applied to superstrata were studied in detail in \cite{Bena:2017geu}. The basic idea is that the 6-dimensional metric ($\ref{dsDoubleFiber}$) is a double circle fibration in the $(v,\psi)$ coordinates, so it is possible to impose coordinate redefinitions that mix the two coordinates into new angular coordinates $(\hat{v},\hat{\psi})$ as
\begin{align}
\frac{\hat{v}}{R} = \mathbf{a} \frac{v}{R}+\mathbf{b} \psi \qquad \text{and} \qquad \hat{\psi} = \mathbf{c} \frac{v}{R}+\mathbf{d}\psi~, \label{SpecTrans}
\end{align} 
with re-identified periodicities
\begin{align}
\hat{v} \cong \hat{v}+2 \pi R \qquad \text{and} \qquad \hat{\psi} \cong \hat{\psi}+4\pi ~. 
\end{align} 
The parameters $(\mathbf{a},\mathbf{b},\mathbf{c},\mathbf{d})$ are required to form an element of SL$(2,\mathbb{Q})$, i.e. $\mathbf{a},\mathbf{b},\mathbf{c},\mathbf{d} \in \mathbb{Q}$ and are constrained by $\mathbf{a}\mathbf{d}-\mathbf{b}\mathbf{c}=1$. The origin of requiring the group to be SL$(2,\mathbb{Q})$ rather than SL$(2,\mathbb{R})$ is so that the coordinate periodicity re-identifications are well defined. The re-identification may modify the presence/absence of orbifold singularities and so the spectral transformation is not necessarily a diffeomorphism, but is closely related. 

The coordinate transformation (\ref{SpecTrans}) alters the GH base, so the BPS equation (\ref{eqZ1Theta2})-(\ref{BPS6D2}) are modified. The rules for how the data $(V,A,\beta,F,\omega,Z_{I},\Theta^{(I)})$ transform under spectral flow to maintain a BPS solution were derived in \cite{Bena:2008wt} and further refined in \cite{Bena:2017geu}. In order to summarize the transformations it is convenient to introduce the auxiliary data $(K_{3},\xi,\mu,\varpi,\nu)$ defined implicitly by 
\begin{align}
\beta= \frac{K_{3}}{V}(d\psi +A)+\xi~, \qquad   \omega = \mu (d\psi +A) + \varpi~,\qquad P= K_{3}\left(\frac{K_{3}}{V}\nu+\mu \right)~. \label{betaomegaPdef}
\end{align}
The transformations for $(V,A,\beta,F,\omega,Z_{I})$ are then given by
\begin{align}
\widehat{V} = \mathbf{d} V - \frac{\mathbf{c}}{R} K_{3}~, \qquad \widehat{K}_{3} = -\mathbf{b}R V+\mathbf{a}K_{3}~, \qquad \hat{\xi} = \mathbf{a}\xi +\mathbf{b}R A~, \qquad \widehat{A}=\frac{\mathbf{c}}{R}\xi+\mathbf{d}A~, \label{SpecVars1}
\end{align}
\begin{align}
\widehat{\varpi}=\varpi~, \qquad \hat{\nu} = \frac{\mathbf{a}\nu K_{3}^{2}+\mathbf{b} R\mu V^{2}}{\widehat{K}_{3}^{2}}~, \qquad \hat{\mu} = \frac{\frac{\mathbf{c}}{R}\nu K_{3}^{2}+\mathbf{d}\mu V^{2}}{\widehat{V}^{2}}~, \qquad \widehat{F} = \frac{\widehat{V}}{V}F -2 \frac{\mathbf{c}}{R} \mu - \frac{\mathbf{c}^{2}}{R^{2}} \frac{P}{\widehat{V}}~, \label{SpecVars2}
\end{align}
and
\begin{align}
\widehat{Z}_{I} = \frac{V}{\widehat{V}} Z_{I}~. \label{SpecZs}
\end{align}

The $\Theta^{(I)}$ transformations are more involved. It is useful to introduce $\lambda^{(I)}$ defined by 
\begin{align}
\Theta^{(I)} = (1+*)\left[(d\psi +A)\wedge \lambda^{(I)} \right]~,
\end{align}
as well as the covariant derivative\footnote{Note that $d_{3}$ is the exterior derivative with respect to the 3D base in (\ref{GHbase}) which is invariant under spectral flow.}
\begin{align}
\mathcal{D} = d_{3} - A \partial_{\psi}-\xi \partial_{v}
\end{align}
which is conveniantly invariant under spectral transformation, i.e. $\mathcal{D}=\widehat{\mathcal{D}}$ with 
\begin{align}
\widehat{\mathcal{D}}=d_{3} - \widehat{A} \partial_{\hat{\psi}}-\hat{\xi} \partial_{\hat{v}}~.
\end{align}
Using these definitions the $\lambda^{(I)}$ transformations can be written as
\begin{align}
\hat{\lambda}^{(1)} = \lambda^{(1)} - \frac{\mathbf{c}}{R} \widehat{\mathcal{D}}\left( \frac{\widehat{Z}_{2}}{V}\right)~, \qquad \hat{\lambda}^{(2)} = \lambda^{(2)} - \frac{\mathbf{c}}{R} \widehat{\mathcal{D}}\left( \frac{\widehat{Z}_{1}}{V}\right)~, \qquad \hat{\lambda}^{(4)} = \lambda^{(4)} - \frac{\mathbf{c}}{R} \widehat{\mathcal{D}}\left( \frac{\widehat{Z}_{4}}{V}\right)~, \label{Speclambdas}
\end{align}
and the $\widehat{\Theta}^{(I)}$ take the form
\begin{align}
\widehat{\Theta}^{(I)} = (1+\hat{*})\left[(d\hat{\psi} +\widehat{A})\wedge \hat{\lambda}^{(I)} \right]~.\label{SpecThetas}
\end{align}

In 5 dimensions a subset of the spectral transformations are gauge transformations, these correspond to keeping $\psi$ fixed, but shifting $v$ by a multiple of $\psi$
\begin{align}
\frac{\hat{v}}{R} = \frac{v}{R}+\mathbf{b} \psi \qquad \text{and} \qquad \hat{\psi} = \psi~. \label{GaugeTrans}
\end{align}
Such a transformation when implemented in equations (\ref{SpecVars1})-(\ref{SpecZs}) and (\ref{Speclambdas})-(\ref{SpecThetas}) leaves the physical data $(Z_{I},Z_{3},\Theta^{I},\Theta^{(3)},\mu,\varpi)$ invariant. A discussion of the general form of these gauge transformations requires a decomposition of the physical data into a set of harmonic functions \cite{Bena:2017geu,Bena:2008wt}. It is important to recognize this is only a gauge transformation in the 5-dimensional setting, in 6-dimensions the metric (\ref{ds6}) depends explicitly on $\beta$ and although $\widehat{\Theta}^{(3)}=\Theta^{(3)}$, $\beta$ transforms as
\begin{align}
\widehat{\beta} = \beta - \mathbf{b} R\, d\psi~.
\end{align}
Thus the transformation (\ref{GaugeTrans}) is physically relevant in 6 dimensions but not in 5 dimensions\footnote{This is why the massless wave equations considered in section \ref{Sect:Sect5 Seperability} depend on $\mathbf{a}$ only in terms with $p$ coefficients}. 

\subsection{6D $\iff$ 5D solutions for single-mode superstrata} \label{SubSec: 6D5D relationship}
In \cite{Bena:2017geu} it was noted that if a 6-dimensional superstrata is independent of $v$ it is simple to reduce the solution to a 5-dimensional solution. However, it was not fully appreciated that there always exists a transformation of the form (\ref{SpecTrans}) that makes $v_{k,m,n}$ of (\ref{moding}) to be independent of $\hat{v}$. This means that for any single-mode superstrata there exists a spectral transformation after which it can be reduced to a 5-dimensional solution. The trade off one makes is that the flow turns the flat $\mathbb{R}^{4}$ base on which the superstrata were first constructed into an ambipolar two centered GH base. This could be anticipated since in 5-dimensions the only non-trivial topology capable of supporting non-singular fluxes is the GH base, whereas the 6-dimensional solutions with a flat base exploit the topology of the $v$ fiber to support non-singular fluxes\footnote{It is this topological dependence on the $v$-fiber that is at the heart of why it is difficult to generalize the results of \cite{Tyukov:2018ypq} to 6-dimensions and find prepotentials for the fluxes.}.  

The asymptotic geometry in 5 dimensions will depend on the net GH charge. If it is zero, as for the 5-dimensional examples in \cite{Bena:2017geu}, then it is $\text{AdS}_{3}\times \mathbb{S}^{2}$. But if the net GH charge is $q\neq 0$, it will be asymptotically $\text{AdS}_{2}\times \mathbb{S}^{3}/\mathbb{Z}_{q}$. The former is appropriate for the microstate geometries of black strings and the latter to those of black holes in 5 dimensions. Since our construction produces both types, we have found the first examples of superstrata that describe the microstates of black holes in 5 dimensions .

To find the spectral transformations (\ref{SpecTrans}) that transform (\ref{moding}) to be $\hat{v}$ independent, it is useful to look at just the parts of the mode (\ref{moding}) that are altered by the spectral transformation (\ref{SpecTrans}), so we define
\begin{align}
\chi_{k,m,n} = (m+n)\frac{v}{R}+\frac{1}{2}(k-2m)\psi~.
\end{align}
Applying the transformation (\ref{SpecTrans}) leads to the new mode dependence \begin{align}
\hat{\chi}_{k,m,n} = \begin{pmatrix}
m+n & \frac{k-2m}{2}
\end{pmatrix} \begin{pmatrix}
\mathbf{d} & -\mathbf{b} \\ -\mathbf{c} & \mathbf{a}
\end{pmatrix} \begin{pmatrix}
\frac{\hat{v}}{R} \\ \hat{\psi}
\end{pmatrix}~.
\end{align}
Demanding $\hat{v}$ independence and fixing $\mathbf{a}\mathbf{d}-\mathbf{b}\mathbf{c}=1$ ensures 
\begin{align}
\mathbf{c} =  \frac{2(m+n)}{\mathbf{e}} \qquad \text{and} \qquad \mathbf{d}=\frac{k-2m}{\mathbf{e}}~, \label{SpecificSF}
\end{align}
with the new mode dependence
\begin{align}
\hat{v}_{k,m,n} = \frac{1}{2}(\mathbf{e}\hat{\psi} - k\phi)~,
\end{align}
where we have defined
\begin{align}
\frac{\mathbf{e}}{2}=\mathbf{a} \left(\frac{k-2m}{2} \right) -\mathbf{b}(m+n)~. \label{edef}
\end{align}
We see that in order to have the correct periodicity in $\hat{\psi}$, it must be that $\mathbf{e}\in \mathbb{Z}$.

Using the standard $(\beta,V)$ of the 6-dimensional superstrata
\begin{align}
\beta = -\frac{R a^{2}}{2 \Sigma}\left(d\phi +\cos 2\theta \, d\psi \right)~, \qquad V = \frac{1}{r_{+}}~, \label{beta and V}
\end{align}
and a spectral transformation of the form (\ref{SpecVars1}) constrained by (\ref{SpecificSF}) and (\ref{edef}) then leads to a two centered ambipolar GH base with 
\begin{align}
\widehat{V} = \frac{q_{-}}{r_{-}} + \frac{q_{+}}{r_{+}}\qquad \text{with} \qquad q_{-} = -\frac{m+n}{\mathbf{e}} \qquad \text{and} \qquad q_{+} = \frac{k-m+n}{\mathbf{e}}~. \label{GHchargesFlowed}
\end{align}

It is now clear that the integers $(\mathbf{e},k)$ control the mode of the flowed solution and $(m,n)$ control the GH charges. If we make the choice to trade $\mathbf{b}$ for $\mathbf{e}$ then there is one remaining degree of freedom, $\mathbf{a}$. This parameter implements a gauge transformation of the form (\ref{GaugeTrans}), as can be seen directly by considering two spectral flows differing by a choice of $\mathbf{a}$. Consider the two transformations
\begin{align*}
\begin{pmatrix}
\hat{v}_{1}/R \\ \hat{\psi}_{1}
\end{pmatrix} = \begin{pmatrix}\mathbf{a}_{1} & \frac{\mathbf{a}_{1}(k-2m)-\mathbf{e}}{2(m+n)}\\ \frac{2(m+n)}{\mathbf{e}}  & \frac{k-2m}{\mathbf{e}}\end{pmatrix} \begin{pmatrix}
v/R \\ \psi
\end{pmatrix}~, \qquad \begin{pmatrix}
\hat{v}_{2}/R \\ \hat{\psi}_{2}
\end{pmatrix} = \begin{pmatrix}\mathbf{a}_{2} & \frac{\mathbf{a}_{2}(k-2m)-\mathbf{e}}{2(m+n)}\\ \frac{2(m+n)}{\mathbf{e}}  & \frac{k-2m}{\mathbf{e}}\end{pmatrix} \begin{pmatrix}
v/R \\ \psi
\end{pmatrix}~,
\end{align*}
computing the differences in the transformed coordinates gives
\begin{align}
\frac{\hat{v}_{2}}{R} = \frac{\hat{v}_{1}}{R} + \frac{\mathbf{e}(\mathbf{a}_{2}-\mathbf{a}_{1})}{2(m+n)} \hat{\psi}_1 \qquad \text{and} \qquad \hat{\psi}_{2} = \hat{\psi}_{1}~,  
\end{align}
which is exactly a gauge transformation of the form (\ref{GaugeTrans}) between $(\hat{v}_{1},\hat{\psi}_{1})$ and $(\hat{v}_{2},\hat{\psi}_{2})$. Hence the role of $\mathbf{a}$ when $(\mathbf{b},\mathbf{c},\mathbf{d})$ are restricted by (\ref{SpecificSF}) and (\ref{edef}) is to implement a gauge transformation in 5-dimensions. We can now enumerate the meaningful degrees of freedom remaining in the 5-dimensional solutions $(\mathbf{e},k,m,n)$, which by (\ref{GHchargesFlowed}) are equivalent to $(\mathbf{e},k,q_{-},q_{+})$. Thus for a given 5-dimensional mode and GH charges there always exists a 6-dimensional superstrata and a spectral transformation that leads to a 5D superstrata with these properties.

Since $(\mathbf{a},\mathbf{b},\mathbf{c},\mathbf{d}) \in \text{SL}(2,\mathbb{Q})$ this process is invertible,  given a BPS solution on a two centered ambipolar GH base in 5-dimensions it can be transformed into a BPS solution on a flat $\mathbb{R}^{4}$ base in 6-dimensions. Thus the identification between 6D$\implies$5D single-mode superstrata is in fact one to one. Given a 5-dimensional solution with a two centered ambipolar GH base it is possible to invert (\ref{6to5data}) to uplift to 6-dimensions, then spectral flow so that the base becomes flat $\mathbb{R}^{4}$. we summarize this result by writing 6D$\iff$5D for single-mode superstrata.

\subsection{6D $\iff$ 5D solutions for multi-mode superstrata?} \label{SubSec: non reduction}
Multi-mode superstrata  are solutions that superpose multiple single-mode superstrata. These solutions involve $(Z_{I},\Theta^{(I)})$ that depend on multiple modes of the form (\ref{moding}), the simplest case being when one considers just two modes labeled by $(k_{1},m_{1},n_{1})$ and $(k_{2},m_{2},n_{2})$. For instance in \cite{Heidmann:2019zws} the $\Theta^{(4)}$ introduced for a simple two-mode solution is
\begin{align}
\Theta^{(4)} = b_{4}\vartheta_{k_{1},m_{1},n_{1}}+b_{5}\vartheta_{k_{2},m_{2},n_{2}} +c_{4}\varphi_{k_{1},m_{1},n_{1}} +c_{5}\varphi_{k_{2},m_{2},n_{2}}~,
\end{align}
where the constants $(b_{4},b_{5},c_{4},c_{5})$ are generically non-zero. The obstruction to reduction is now clear, this flux depends on both $(v_{k_{1},m_{1},n_{1}},v_{k_{2},m_{2},n_{2}})$, a spectral transformation will only remove the $v$-dependence if the modes are \textit{parallel} $(m_{1}+n_{1},k_{1}-2m_{1})\propto (m_{2}+n_{2},k_{2}-2m_{2})$. Generalization to more modes is immediate.

%%%%%%%%%%%%%%%%%%%%%%%%%%%%%%%%%%%%%
\section{Relations amongst superstrata families} 
\label{Sect:Sec4 relations amongst superstrata}
%%%%%%%%%%%%%%%%%%%%%%%%%%%%%%%%%%%%%

This section highlights several relationships amongst superstrata families that were either not known or not highlighted in the current literature. In particular it is shown that after spectral transformation and reduction to 5 dimensions the $(k,m,n)$ and $(k,k-m,n)$ families are equivalent.  

\subsection{Equivalence of 5-dimensional solutions related by signs} \label{Sub sec: signs}
Some simple observations about the structure of the 5D BPS equations (\ref{5D BPS Z1eq})-(\ref{5DBPSfinal}) can by made be summarizing the data upon which it depends
\begin{align}
(Z_{I},Z_{3},\Theta^{(I)},\Theta^{(3)},\omega)~,
\end{align}
and altering some signs. A couple of ``new" solutions can be found by defining the new data $(\widetilde{Z}_{I},\widetilde{Z}_{3},\widetilde{\Theta}^{(I)},\widetilde{\Theta}^{(3)},\widetilde{\omega})$ by either of the following:
\begin{align}
(\widetilde{Z}_{I},\widetilde{Z}_{3},\widetilde{\Theta}^{(I)},\widetilde{\Theta}^{(3)},\widetilde{\omega})&= (-Z_{I},Z_{3},-\Theta^{(I)},\Theta^{(3)},\omega)~, \label{Trans1}\\
(\widetilde{Z}_{I},\widetilde{Z}_{3},\widetilde{\Theta}^{(I)},\widetilde{\Theta}^{(3)},\widetilde{\omega})&= (Z_{I},Z_{3},-\Theta^{(I)},-\Theta^{(3)},-\omega) \label{Trans2}~.
\end{align}
The first of these transformations (\ref{Trans1}) corresponds to a trivial redefinition
\begin{align}
(\widetilde{Q}_{1},\widetilde{Q}_{5},\widetilde{b}_{4},\widetilde{c}_{4}) = (-Q_{1},-Q_{5},-b_{4},-c_{4})~.
\end{align}
The second transformation (\ref{Trans2}) is more subtle, looking back at the 5-dimensional geometry (\ref{ds5}) we see that if one also reverses time $\tilde{t}=-t$ then the geometry is unchanged. If one considers that the $Z_{I}$ control the electric charge and the $\Theta^{(I)}$ the magnetic charge, then these two solutions are indeed just identified by time reversal, and thus equivalent. 

If we look closely at the spectral transformations of section (\ref{SubSec: spectral flow}) we discover a third transformation. Consider the spectral flow that redefines $(\hat{v},\hat{\psi})=(-v,-\psi)$ using the SL$(2,\mathbb{Q})$ transformation
\begin{align}
(\mathbf{a},\mathbf{b},\mathbf{c},\mathbf{d}) = (-1,0,0,-1) \label{Trans3}~.
\end{align}
Under this transformation
\begin{align}
(\widehat{Z}_{I},\widehat{Z}_{3},\widehat{\Theta}^{(I)},\widehat{\Theta}^{(3)},\widehat{\omega})= (-Z_{I},-Z_{3},-\Theta^{(I)},-\Theta^{(3)},\omega)\qquad \text{and} \qquad  \widehat{ds}_{4}^{2}(\mathcal{B}) = -ds_{4}^{2}(\mathcal{B})~,
\end{align}
where it is understood that if there is any functional dependence on $\psi$ in the data it must be replaced by $\hat{\psi}=-\psi$. This relabeling does not obviously lead to a new soltion of the BPS equations, but since the spectral transformation that produces it requires no alterations of the identifications on the $(v,\psi)$ circles, we conclude it is identical to the solution before the transformation was performed.

\subsection{Relating the $(k,m,n)$ and $(k,k-m,n)$ superstrata}
Based on the results of the previous subsection it is worth considering if any of the 6-dimensional superstrata when reduced to 5 dimensions lead to the same solution. Consider two families $(k_{1},m_{1},n_{1})$ and $(k_{2},m_{2},n_{2})$, if they are to posses the same mode dependence after spectral flow, the same $\mathbf{e}$ must be used in each flow and $k_{1}=k_{2}\equiv k$ must be fixed. Consider the situation when the two 5-dimensional GH bases are related by
\begin{align}
-q_{-(k,m_{1},n_{1})}=q_{+(k,m_{2},n_{2})} \qquad \text{and}\qquad -q_{-(k,m_{2},n_{2})}=q_{+(k,m_{1},n_{1})} ~,
\end{align}
which from (\ref{GHchargesFlowed}) fixes
\begin{align}
m_{2}=k-m_{1} \qquad \text{and} \qquad n_{1} =n_{2}~.
\end{align}
The $(Z_{I},\beta,ds_{4}^{2}(\mathcal{B}))$ for the $(k,m,n)$ family after spectral transformation to remove $\hat{v}$ dependence are given by:
\begin{align}
Z_{1} &= \frac{2\mathbf{e}}{\Upsilon_{k,m,n}}\left(Q_{1} + \frac{b_{4}^{2}R^{2}}{Q_{5}} \Delta_{2k,2m,2n}\cos \hat{v}_{2k,2m,2n} \right) \label{ZZZZ1} ~,\\
Z_{2}&= \frac{2\mathbf{e}Q_{5}}{\Upsilon_{k,m,n}}~, \\
Z_{4}&= \frac{2\sqrt{2}\mathbf{e}R}{\Upsilon_{k,m,n}} \Delta_{k,m,n} \cos \hat{v}_{k,m,n} ~,\\
\beta &= \frac{\mathbf{e}R\left[(\mathbf{e}-\mathbf{a}k+2\mathbf{a}m )(a^{2}+2r^{2})+a^{2}(\mathbf{e}-\mathbf{a}k-2\mathbf{a}n)\cos 2\theta\right]}{2(m+n)\Upsilon_{k,m,n}} \, d\psi -\frac{a^{2}\mathbf{e}R}{\Upsilon_{k,m,n}} \, d\phi~,
\end{align}
where
\begin{align}
\Upsilon_{k,m,n}= (k-2m)(a^{2}+2r^{2})+a^{2}(k+2n)\cos 2\theta~.
\end{align}
The GH base is of the form (\ref{GHbase}) with $V$ is given by the $\widehat{V}$ of (\ref{GHchargesFlowed}) and 
\begin{align}
A&=\frac{8 r^2 \left(a^2+r^2\right)   (k-2 m)\cos 2 \theta -2 a^2 \left(a^2+2 r^2\right)  (k+2 n) \sin ^2 2 \theta }{8\mathbf{e}\Sigma \Lambda} d\phi ~. \label{AAAAA1}
\end{align}

Applying the transformations $m\to k-m$,  $\theta\to\frac{\pi}{2}-\theta$, $\phi\to - \phi$, $\psi\to-\psi$, $\mathbf{a}\to\frac{1}{m+n}(\mathbf{e}-\mathbf{a}(k-m+n))$ to the data outlined above and labeling the transformed quantities by tildes gives 
\begin{align}
(\widetilde{Z}_{1},\widetilde{Z}_{2},\widetilde{Z}_{4},\tilde{\beta})=(-Z_{1},-Z_{2},-Z_{4},\beta) \qquad \text{and} \qquad \widetilde{ds}_{4}^{2}(\mathcal{B})=-ds_{4}^{2}(\mathcal{B})~.
\end{align}
The form of the BPS equations then fixes $(\Theta^{(I)},F,\omega)$ and implies the full identification
\begin{align}
(\widetilde{Z}_{I},\widetilde{Z}_{3},\widetilde{\Theta}^{(I)},\widetilde{\Theta}^{(3)},\widetilde{\omega})&=(-Z_{I},-Z_{3},\Theta^{(I)},\Theta^{(3)},-\omega)~, \notag  \\
\widetilde{ds}_{4}^{2}(\mathcal{B})&= -ds_{4}^{2}(\mathcal{B})~. \label{trans6}
\end{align}
Referring to section \ref{Sub sec: signs} we see that this corresponds to a spectral transformation of the form (\ref{Trans3}) followed by a transformation of the form (\ref{Trans2}), thus the $(k,m,n)$ and $(k,k-m,n)$ families are equivelent when reduced 5-dimensions.

\subsection{The $(1,0,n)$ and $(1,1,n)$ original superstrata} \label{SubSec: 11n}
The relationship of the previous subsection can be explicitly demonstrated for the $(1,0,n)$ and $(1,1,n)$ original superstrata families. The $(F,\omega)$ for the $(1,0,n)$ family were already known in closed form \cite{Bena:2016ypk}, while it is possible to compute the closed form for the $(1,1,n)$ family:
\begin{align}
F_{1,0,n} &= \frac{b^{2}}{a^{2}}(\Gamma^{n}-1)~, \qquad ~~ \omega_{1,0,n} = \omega_{0}+ \frac{b^{4}}{b_{4}^{2}} \frac{R}{2\Sigma} \left((\Gamma^{n}-1)\sin^{2}\theta \right)(d\phi - d\psi)~, \label{10n Sol}\\
F_{1,1,n}&= \frac{b^{2}}{a^{2}} (\Gamma^{n+1}-1) ~, \qquad \omega_{1,1,n} = \omega_{0} + \frac{b^{4}}{b_{4}^{2}}\frac{R}{2\Sigma} \left[ \Gamma^{n+1}\cos^{2}\theta \,(d\phi+d\psi)-\sin^{2}\theta \, (d\phi-d\psi) \right]~. \label{11n Sol}
\end{align}
It is interesting that the $F$ and $\omega$ prior to spectral flow are not related in an obvious way.

%%%%%%%%%%%%%%%%%%%%%%%%%%%%%%%%%%%%%
\section{Separability of wave equations in 5 and 6 dimensions} 
\label{Sect:Sect5 Seperability}
%%%%%%%%%%%%%%%%%%%%%%%%%%%%%%%%%%%%%
This section studies the massless wave equations for various superstrata in depth. The results of a search to find superstrata families with separable massless wave equations (SMWEs) in either 5 or 6 dimensions is summarized.

\subsection{General structure of wave equation for axially symmetric BPS solutions} \label{SubSec: wave equation structure}
For superstrata defined on a flat $\mathbb{R}^{4}$ or two center GH bases, the data appearing in (\ref{ds6}) and ({\ref{ds5}}) are axially symmetric and so have the functional dependence  
\begin{align}
V=V(r,\theta)~, \qquad P=P(r,\theta)~, \qquad F=-Z_{3}=F(r,\theta)~, \qquad A=A_{\phi}(r,\theta)\, d\phi~,\label{rest1}
\end{align}
\begin{align}
\omega= \omega_{\phi}(r,\theta) \, d\phi + \omega_{\psi}(r,\theta)~d\psi \qquad \text{and} \qquad \beta =\beta_{\phi}(r,\theta)\, d\phi + \beta_{\psi}(r,\theta)\, d\psi~.  \label{rest2}
\end{align}
The massive wave equation is given by 
\begin{align}
\frac{1}{\sqrt{-g}} \partial_{\mu} \left( \sqrt{-g}g^{\mu\nu}\partial_{\nu}\Phi \right) = M^{2} \Phi~, \label{MassiveWE}
\end{align}
where $g$ denotes the determinate of $g_{\mu\nu}$ which is either (\ref{ds6}) or (\ref{ds5}) depending on whether it is in 6 or 5 dimensions. In 6 dimensions we utilize the periodicity of the $(v,\phi,\psi)$ coordinates and independence of $u$, so look for solutions of the separable form
\begin{align}
\Phi = K(r) S(\theta) e^{i\left(\frac{w}{R} u +\frac{p}{R}v+q_{1}\phi +q_{2}\psi \right)}~, \label{PhiAnsatz}
\end{align} 
where $(w,p,q_{1},q_{2})$ are constants, the 5-dimensional form is obtained by setting $p=0$. The massive wave equation (\ref{MassiveWE}) can then be written as
\begin{align}
 \frac{1}{r}\partial_{r}\left(r(a^{2}+r^{2})\partial_{r}K \right) + \frac{1}{\sin 2\theta} \partial_{\theta} \left( \sin 2\theta \, \partial_{\theta} S \right) + G^{(i)}_{1}(r,\theta) = \frac{M^{2} G^{(i)}_{2}(r,\theta)}{KS}~, \label{WEgen}
\end{align}
where $i\in \{5,6\}$ indexes the 5 or 6-dimensional version. Direct computation gives
\begin{align}
G_{2}^{(6)}=\frac{\Sigma \Lambda}{4} V \sqrt{P}  \qquad \text{and} \qquad G_{2}^{(5)} = \frac{\Sigma \Lambda}{4} V \left(-F P \right)^{1/3}~,
\end{align}
looking at the form of $\Sigma\Lambda$ in (\ref{SigmaLambda}), it is obvious that these terms destroy separability. However, when $M=0$ separability will depend solely on the form of $G_{1}^{(i)}(r,\theta)$. In 6 dimensions 
\begin{align}
G_{1}^{(6)} &= -\frac{\Lambda \Sigma}{R^{2}} \left\lbrace \frac{4\Gamma}{r^{4}}\left[q_{1}R-p \beta_{\phi}-w \omega_{\psi}+A_{\phi}(-q_{2}R +p \beta_{\psi}+w \omega_{\psi}) \right]^{2} \right. \\
& \qquad\qquad\qquad\qquad \left. + \frac{V}{4}  \left[w(-2p+w F)P+V(-q_{2}R+p \beta_{\psi}+w \omega_{\psi})^{2} \right] \right\rbrace ~.\label{6DG}
\end{align}
It is convenient to expand $G_{1}^{(6)}(r,\theta)$ in the form 
\begin{align}
G_{1}^{(6)}(r,\theta) =\frac{1}{2} \sum_{x_{1},x_{2}\in \mathcal{S}}x_{1}x_{2} G_{x_{1}x_{2}}(r,\theta) ~,\label{Gscheme}
\end{align}
where $\mathcal{S}=\left\lbrace w,p,q_{1},q_{2} \right\rbrace$ and $G_{x_{1},x_{2}}$ are functions. For future reference the form of the $G_{x_{1}x_{2}}(r,\theta)$ are summarized in a table in appendix \ref{App: 1}, where we have introduced the function
\begin{align}
\Gamma = \frac{r^{2}}{a^{2}+r^{2}}~.
\end{align}

The convenience of introducing things this way is that simply dropping the terms proportional to $p$ and setting $F=-Z_{3}$ gives the 5-dimensional result:
\begin{align}
G_{1}^{(5)}(r,\theta)=\left. G_{1}^{(6)}(r,\theta) \right|_{p= 0,F= -Z_{3}}~. \label{G6Dto5D}
\end{align}
This makes the tables appearing in the appendices useful, the full tables give the 6-dimensional result, omitting the last four rows gives the 5-dimensional result.

The massless wave equations will be separable if every $G_{x_{1}x_{2}}(r,\theta)$ term splits into a sum of a function of $r$ and a function of $\theta$ alone. Given (\ref{G6Dto5D}) there is also the possibility for the 6-dimensional wave equation to be non-separable whilst the 5-dimensional one is, if non-separable terms only appear in terms with a factor of $p$. Section \ref{SubSec: 21n sep} shows that this occurs for the $(2,1,n)$ original family.

\subsection{Separability of $(1,0,n)$ and $(1,1,n)$ original superstata} \label{SubSec: 10n 11n wave}
In \cite{Bena:2017upb} it was shown that the $(1,0,n)$ family have a SMWEs in 6-dimensions. Since spectral transformations leave $(r,\theta)$ inert, performing such a transformation should not effect the seperability of (\ref{Gscheme}). We performed the spectral transformation procedure of section \ref{Sect:Sec3 relating 5D 6D superstrata}, to both the $(1,0,n)$ and $(1,1,n)$ families, thus removing any $v$-dependence in the solutions. The resulting $G_{x_{1}x_{2}}$ for the wave equations of these families are summarized in appendix \ref{App: 2}, table \ref{10n table}. The 5-dimensional result is given again by applying (\ref{G6Dto5D}) and dropping the last 4 rows of the table. 
 
Table \ref{10n table} possesses some interesting features:
\begin{itemize}
\item In addition to the $(1,0,n)$ family having a SMWEs in 6 dimensions, the $(1,1,n)$ family has SMWEs as well.
\item Both families have SMWEs in 5 dimensions. 
\item The remaining 6-dimensional spectral transformation parameters $(\mathbf{a},\mathbf{e})$ alter the form of the wave equations substantially, whilst maintaining separability. It is possible to set either 
\begin{align}
\mathbf{a}(1+2n)-\mathbf{e}=0 \qquad \text{or} \qquad \mathbf{a}-\mathbf{e}=0~,
\end{align}
and simplify either the $r$ or $\theta$ dependent parts of the wave equation.
\item Redefining $\tilde{\theta}=\frac{\pi}{2}-\theta$ for one of the families, we see that the 5-dimensional wave equations become identical. As was required by the identification of these solutions in section \ref{SubSec: 11n}.
\item The spectral transformation parameter $\mathbf{a}$ does not appear in any of the 5-dimensional terms since it is a gauge transformation (see section \ref{SubSec: 6D5D relationship}), but it does appear in the 6-dimensional terms. 
\end{itemize}

\subsection{Separability of $(2,1,n)$ superstata} \label{SubSec: 21n sep}
In \cite{Bena:2017upb} it was shown how the 6-dimensional $(2,1,n)$ original superstrata family has SMWEs so long as the momentum on the GH fiber direction $\psi$ vanishes\footnote{In \cite{Bena:2017upb} it was not acknowledged explicitly that this was the GH fiber direction, the choice of coordinates there obscured this fact.}. Using the spectral transformations of section \ref{SubSec: 6D5D relationship} this becomes a constraint requiring the momentum on the $v$-circle to vanish, the 5-dimensional reduction should thus have SMWEs. In contrast the supercharged $(2,1,n)$ family has SMWEs 6 dimensions alaready \cite{Heidmann:2019zws}. 

To aid in the presentation of the wave equations for the $(2,1,n)$ family we schematically break up the $G_{x_{1}x_{2}}(r,\theta)$ of (\ref{WEgen}) into pieces distinguished by their dependence on $b$ or $c$:
\begin{align}
G_{x_{1}x_{2}}=  G^{(0)}_{x_{1}x_{2}}+G^{(b)}_{x_{1}x_{2}}+G^{(c)}_{x_{1}x_{2}}+G^{(bc)}_{x_{1}x_{2}}~,
\end{align}
where we define
\begin{align}
G^{(0)}_{x_{1}x_{2}}&=\left.G_{x_{1}x_{2}}\right|_{b=0,c=0}~,\qquad\qquad G^{(b)}_{x_{1}x_{2}}=\left.G_{x_{1}x_{2}}\right|_{c=0}-G^{(0)}_{x_{1}x_{2}}~,\\
\qquad G^{(c)}_{x_{1}x_{2}}&=\left.G_{x_{1}x_{2}}\right|_{b=0}-G^{(0)}_{x_{1}x_{2}} ~,\qquad\, G^{(bc)}_{x_{1}x_{2}}=G_{x_{1}x_{2}}-\left(G^{(0)}_{x_{1}x_{2}}+G^{(b)}_{x_{1}x_{2}}+G^{(c)}_{x_{1}x_{2}} \right)~.
\end{align}
Thus the original superstrata result is given by the $(G^{(0)}_{x_{1}x_{2}},G^{(b)}_{x_{1}x_{2}})$ terms,the supercharged result by the $(G^{(0)}_{x_{1}x_{2}},G^{(c)}_{x_{1}x_{2}})$ terms, and the hybrid result by the full $G_{x_{1}x_{2}}$. 

The result of the wave equation analysis are presented in appendix \ref{App 3}, table \ref{21n table}. It has several interesting features:
\begin{itemize}
\item As highlighted in \cite{Heidmann:2019zws} the supercharged $(2,1,n)$ family have SMWEs in 6 dimensions.
\item The original $(2,1,n)$ family fail to have SMWEs in 6 dimensions due to the term
\begin{align}
G_{pw}^{(b)}(r,\theta)=\frac{b^2 \Gamma  (2 \mathbf{a} (n+1)-\mathbf{e}) \left(a^2 \Gamma ^{n+1}+2 r^2\right)}{2 (n+1) r^4}+\frac{a^2 b^2 \mathbf{e}  \Gamma ^{n+2} \cos 2 \theta }{r^4}~.
\end{align}
\item Both the original and supercharged flavors have SMWEs in 5 dimensions.
\item Unlike the $(1,0,n)$ and $(1,1,n)$ families there is now only one obvious choice for fixing $(\mathbf{a},\mathbf{e})$ to simplify the 6-dimensional wave equations 
\begin{align}
2\mathbf{a}(n+1)-\mathbf{e}=0~.
\end{align}
\end{itemize}

There are two non vanishing $G^{(bc)}_{x_{1}x_{2}}$ terms for the hybrid $(2,1,n)$ family. The term 
\begin{align}
G^{(bc)}_{pw}(r)=\frac{ \Gamma ^2 \mathbf{e} \left(\Gamma ^n \left(a^4 (n+1) (n+2)+2 a^2 (n+2) r^2+2 r^4\right)-2 \left(a^2+r^2\right)^2\right)}{a^2 \sqrt{n} (n+1) \sqrt{n+2} r^4}~,
\end{align}
which is removed in the 5-dimensional reduction. As well as 
\begin{align}
G^{(bc)}_{wq_{1}}(r,\theta) = \left(\frac{2 b c  \left(\Gamma ^{n+2} \left(a^4 (n+1) (n+2)+2 a^2 (n+2) r^2+2 r^4\right)-2 r^4\right)}{a^2 \sqrt{n (n+2)} r^4}\right) \cos 2\theta~,
\end{align}
which is non-separable, but is removed upon setting $q_{1}=0$. Thus the hybrid solutions don't have SMWEs in 6 dimensions due to the $G^{(b)}_{pw}$ and $G^{(bc)}_{wq_{1}}$ terms, but they will have SMWEs in 5 dimensions if one sets $q_{1}=0$. This also implies the null geodesics with no motion in the $\phi$-direction can be solved for analytically for the hybrid $(2,1,n)$ family in 5 dimensions.

\section{Prepotentials} \label{Sect 3.5: prepotentials}
The $(\Theta^{(I)},\Theta^{(3)})$ fluxes in 5 dimensions can be derived from prepotential functions $(\Phi^{(I)},\Phi^{(3)})$ on the GH base \cite{Tyukov:2018ypq}. Once the possible prepotentials are characterized and it is understood what moduli of the hyper-K\"{a}hler base the corresponding fluxes control, the $(Z_{I},Z_{3})$ are simply derived from the prepotenitals without the need to solve any differential equations. In this section we summarize the prepotential construction, uplift the construction to 6-dimensions and compute the prepotentials for the superstrata fluxes (\ref{thetatilde}) and (\ref{thetahat}) with arbitrary $(k,m,n)$. Previously, the the only known examples of prepotentials were for the original superstrata fluxes (\ref{thetatilde}) for $k=2m$ in 5 dimensions.

\subsection{Prepotentials in 5-dimensions} \label{SubSec: 5D prepotentials}
In \cite{Tyukov:2018ypq} it was shown that in 5-dimensions the $\Theta^{(I)}$ can be derived from harmonic functions on the GH base known as prepotentials $\Phi^{(I)}$. The construction is
\begin{align}
\Theta^{(I)} = d \left(\tensor{J}{_{\mu}^{\nu}}\partial_{\nu} \Phi^{(I)} \, dx^{\mu} \right)~,
\end{align} 
where $J$ is the complex structure. For axisymmetric multi-centered GH bases the canonical complex structure given
\begin{align}
J=(d\psi +A) \wedge dy^{3} -V\, dy^{1}\wedge dy^{2}~,
\end{align}
where the GH charges are coincident on the $y^{3}$ axis and $x^{\mu}\in (\psi,r,\theta,\phi)$. 

The $(Z_{I},Z_{3})$ that solve the BPS equations (\ref{5D BPS Z1eq})-(\ref{5D BPS Z4eq}) can in principle be found without solving any differential equations. Given any harmonic $(1,1)$ form $\Theta$, a perturbation of a Ricci-flat K\"{a}hler manifold with metric $g_{\mu\nu}$ such that it stays Ricci-flat and K\"{a}hler is given by
\begin{align}
\delta g_{\mu\nu} = \frac{1}{2}\left(\tensor{J}{_\mu^\rho}\Theta_{\rho\nu} +\tensor{J}{_\nu^\rho}\Theta_{\rho\mu}  \right)~.
\end{align}
If there is a family of Ricci-flat K\"{a}hler manifolds with parameter $a$, such as the two centered GH bases with (\ref{GHchargesFlowed}), then one might expect $\partial_{a}g_{\mu\nu}=\delta g_{\mu\nu}$. This is true modulo an infinitesimal coordinate change $x^{\mu}\to x^{\mu}+Y^{\mu}_{(a)}$. This vector field $Y^{\mu}_{(a)}$ can be fixed by introducing the covariant derivative
\begin{align}
\mathcal{D}_{a} \equiv \partial_{a} + \mathcal{L}_{Y_{(a)}}
\end{align}
and demanding 
\begin{align}
\mathcal{D}_{a}J=\Theta~, \qquad \mathcal{D}_{a}g_{\mu\nu} =\frac{1}{2}\left(\tensor{J}{_\mu^\rho}\Theta_{\rho\nu} +\tensor{J}{_\nu^\rho}\Theta_{\rho\mu}  \right)~,
\end{align}
where $ \mathcal{L}_{Y_{(a)}}$ is the Lie derivative with respect to the vector field $Y^{\mu}_{(a)}$. 

If there is another harmonic function $\widehat{\Theta}$ with flux $\widehat{\Theta} =  d \left(\tensor{J}{_{\mu}^{\nu}}\partial_{\nu} \Phi \, dx^{\mu} \right)$ then
\begin{align}
\nabla^{2} \left( \mathcal{D}_{a}\widehat{\Phi}\right) = \star \left( \Theta \wedge \widehat{\Theta}\right)~. \label{Zprepotentials}
\end{align} 
In principle this allows gives the $Z_{I}$ that solve (\ref{5D BPS Z1eq})-(\ref{5D BPS Z4eq}) directly from the $\Phi^{(I)}$. However, in practice it is only known how to construct $\mathcal{D}_{a}$ for $\Theta$ of the form $\Theta=d\beta$ as appears in (\ref{beta and V}). It is necessary to understand what modulus of the GH base the $\Theta$ controls, for $d\beta$ it is known to be the spacing between the GH charges parametrized by $a$. The modulus modulus controlled by the superstrata fluxes (\ref{Theta1})-(\ref{Theta4}) after 5-dimensional reduction are unknown.

\subsection{Prepotentials in 6-dimensions} \label{SubSec: 6D prepotentials}
To motivate/derive the results of \cite{Tyukov:2018ypq} it was important that the $\Theta^{(I)}$ were supported solely by the homology on a hyper-K\"{a}hler base. For generic 6-dimensional supertrata this is certainly not the case since the canonical GH base used is just flat $\mathbb{R}^{4}$, indeed the non-trivial homology is due to the the pinching off of the $v$-circle with this base. However, using the spectral transformations of section \ref{Sect:Sec3 relating 5D 6D superstrata} to remove the $v$-dependence of the $\Theta^{(I)}$, ensuring they are supported solely on the homology on the GH base again, even in 6 dimensions. Thus in order to derive the prepotentials for (\ref{thetatilde}) and (\ref{thetahat}) the appropriate spectral flow must be made. Using hats to represent quantities after spectral flow we define
\begin{align}
\widehat{J}_{k,m,n}=(d\hat{\psi} +\hat{A}) \wedge dy^{3} -\widehat{V}\, dy^{1}\wedge dy^{2}~, \qquad \widehat{d}_{k,m,n} \Phi = (\partial_{\hat{\psi}}\Phi) \, d\hat{\psi} + d_{3} \Phi~.
\end{align}
With $(\mathbf{a},\mathbf{b},\mathbf{c},\mathbf{d})$ satisfying (\ref{SpecificSF}) and (\ref{edef}) direct computation gives
\begin{align}
\widehat{J}_{k,m,n} = \frac{r \cos 2\theta}{2} d\hat{\psi} \wedge dr + \frac{(a^{2}+2r^{2})\sin 2\theta}{4} d\theta  \wedge d\hat{\psi}  + \frac{(k-2m)r}{2\mathbf{e}}d\phi \wedge d r +\frac{a^{2}(k+2n)\sin 2\theta}{4 \mathbf{e}} d\phi \wedge d\theta ~.
\end{align}
Noting the form of equations (\ref{Speclambdas}), the $\widehat{\Theta}^{(I)}$ will be of the form
\begin{align}
\widehat{\Theta}^{(1)} &= Q_{5} \kappa ~, \\
\widehat{\Theta}^{(2)} &=Q_{1} \kappa +\frac{R}{\sqrt{2} Q_{5}}\left( b_{1} \widehat{\vartheta}_{2k,2m,2n} + c_{2} \widehat{\varphi}_{2k,2m,2n} \right)~,\\
\widehat{\Theta}^{(4)} &=  b_{4} \widehat{\vartheta}_{k,m,n} + c_{4} \widehat{\varphi}_{k,m,n}~,
\end{align}
where $(\widehat{\vartheta}_{k,m,n},\widehat{\varphi}_{k,m,n})$ are the spectral flowed versions of $(\vartheta_{k,m,n},\varphi_{k,m,n})$ and computation gives 
\begin{align*}
\kappa &= \frac{8 a^2  (m+n) (k-m+n) }{R \left(\left(a^2+2 r^2\right) (k-2 m)+a^2 \cos (2 \theta ) (k+2 n)\right)^2}\widehat{\mathcal{J}}_{k,m,n}~, \\ 
\widehat{\mathcal{J}}_{k,m,n}&= \frac{r \cos 2\theta}{2} d\hat{\psi} \wedge dr - \frac{(a^{2}+2r^{2})\sin 2\theta}{4} d\theta  \wedge d\hat{\psi}  + \frac{(k-2m)r}{2\mathbf{e}}d\phi \wedge d r -\frac{a^{2}(k+2n)\sin 2\theta}{4 \mathbf{e}} d\phi \wedge d\theta ~.
\end{align*} 

Written in terms of a self dual two form basis the expressions for the $(\widehat{\vartheta}_{k,m,n},\widehat{\varphi}_{k,m,n})$ are more complicated than those of (\ref{thetatilde}) and (\ref{thetahat}) prior to the flow. However, using the prepotential prescriptions 
\begin{align}
\widehat{\vartheta}_{k,m,n}  = \widehat{d}_{k,m,n} \left( (\widehat{J}_{k,m,n})_{\mu}^{~~\nu}\partial_{\nu} \widehat{\Phi}^{(\vartheta)}  dx^{\mu}\right) \qquad \text{and} \qquad \widehat{\varphi}_{k,m,n}  = \widehat{d}_{k,m,n} \left( (\widehat{J}_{k,m,n})_{\mu}^{~~\nu}\partial_{\nu} \widehat{\Phi}^{(\varphi)}  dx^{\mu}\right)~,
\end{align}
they can be summarized as
\begin{align}
\widehat{\Phi}^{(\vartheta)} &=C_{1} \frac{\cos \hat{v}_{k,m,n}}{\Delta_{k,m,n}} - \frac{\Delta_{k,m,n}\cos \hat{v}_{k,m,n} }{\sqrt{2}a^{2}k\mathbf{e}} \left[  \left(a^2+r^2\right) (k-2 m) \, _2F_1\left(1,1-k;n+1;-\frac{r^2}{a^2}\right) \right. \notag \\
& \qquad\qquad \qquad  \qquad \qquad \qquad   + a^{2}\left(m+n-(k+2n) \, _2F_1\left(1,k+1;m+1;\cos ^2 \theta \right) \, \sin^{2}\theta\right) \Big]~, \\
\widehat{\Phi}^{(\varphi)} &= C_{2} \frac{\cos \hat{v}_{k,m,n}}{\Delta_{k,m,n}} \notag\\
& \qquad -\frac{\Delta_{k,m,n}\cos \hat{v}_{k,m,n}}{\sqrt{2}a^{2}(k-m)(k+n)mn \mathbf{e}} \left[ m(k-m)(k+2n)(a^{2}+r^{2})  \, _2F_1\left(1,1-k;n+1;-\frac{r^2}{a^2}\right)  \right. \notag \\
&   \qquad \qquad    + a^{2}n \left( m(m+n)+(k-2m)(k+n)\, _2F_1\left(1,k+1;m+1;\cos ^2\theta \right) \sin^{2}\theta \right) \Big]~,
\end{align}
where $C_{1}$ and $C_{2}$ are constants. These prepotentials can be used in the canonical flat $\mathbb{R}^{4}$ base by performing the required inverted spectral transformation. This will also work for multi-mode solutions, but different flows will be needed for individual modes. 

The $\widehat{\Phi}^{(\vartheta)}$ prepotentials are for the original supertrata fluxes, while $\widehat{\Phi}^{(\varphi)}$ correspond to the supercharged potentials. Previously only the $k=2m$ original superstrata prepotentials were known. By extending to all $(k,m,n)$, as well as the supercharged case, we see that the structure is far richer. For instance it was previously unknown that harmonic prepotentials on 2 centered GH bases could be constructed using $ _2F_1\left(1,1-k;n+1;-\frac{r^2}{a^2}\right)$. Such terms appear in the supercharged flavor exactly when $k=2m$, as well as in the original flavor when $k\neq 2m$. 

It is hoped that a mathematical framework can be developed based on functional analysis on ambi-polar GH bases, that might allow one to construct all prepotentials relavent for superstrata. Additionally it would be extremely useful to determine the moduli the corresponding fluxes control and integrate them to produce new hyper-K\"{a}hler bases. It is possible the prepotentials displayed above could aid in this program.

%%%%%%%%%%%%%%%%%%%%%%%%
\section{Discussion, conclusion and outlook}
\label{Sect:Sect6 Discussion}
%%%%%%%%%%%%%%%%%%%%%%%%
It was shown how to transform any single-mode superstrata in 6 dimensions to become independent of the $v$ coordinate using a spectral transformations. This alters the base from flat $\mathbb{R}^{4}$ to ambipolar 2 centered GH, trading the three mode numbers $(k,m,n)$ for the new mode numbers $(\mathbf{e},k)$ and GH charges $(q_{-},q_{+})$. Once this flow has been made it is straightforward  to reduce to a 5-dimensional solution, in fact the $(\mathbf{e},k,q_{-},q_{+})$ are sufficient to parametrize the most general two centered 5-dimensional superstrata. These 5-dimensional solutions include both asymptotically $\text{AdS}_{2}\times \mathbb{S}^{3}$ and $\text{AdS}_{3}\times \mathbb{S}^{2}$ examples. Corresponding to microstate geometries of black strings and black holes respectively. Examples of the black string microstate geometries had been considered in \cite{Bena:2017geu}, while the black hole microstate geometries are new and correspond to having non-zero net GH charge. 
 
The dimensional reduction will fail if one cannot find a spectral transformation that removes the dependence of the data $(Z_{I},Z_{3},\Theta^{(I)},\Theta^{(3)},F,\omega)$ on $v$. This occurs for multi-mode superstrata, unless the distinct modes are arranged to be parallel. Additionally it was shown that the $(k,m,n)$ and $(k,k-m,n)$ superstrata both reduce to the same 5-dimensional solutions. There should be more states in the 6-dimensional as the added dimension allows for a larger event horizon of the black hole being approximated. Thus it's interesting that there are both 6-dimensional solutions that don't reduce to 5 dimensions and different 6-dimensional solutions that reduce to the same 5-dimensional solutions.  

A search for superstrata with SMWEs was conducted. Previously it was known that the original $(1,0,n)$ and supercharged $(2,1,n)$ superstrata in 6-dimensions had SMWEs. These families were spectrally transformed into the form appropriate for dimensional reduction, the remaining spectral transformation parameters $(\mathbf{a},\mathbf{e})$ then indexed families of distinct 6-dimensional 2 centered superstrata with different SMWEs. The 5-dimensional reductions of these solutions were necessarily also have SMWEs. The original $(2,1,n)$ superstrata solutions were known to be non-separable in 6-dimensions, we showed that in 5-dimensions the obstruction is removed and SMWEs are produced. If the momentum around the $\phi$-circle vanishes the hybrid $(2,1,n)$ family also have SMWEs in 5 dimensions. We also showed that the $(1,1,n)$ original superstrata have SWMEs in 6 dimensions. 

Separability of a massless wave equation implies the existence of a conformal killing tensor. Since the 5-dimensional geometries are independent of $(t,\phi,\psi)$, there are enough conserved quantities in our examples to solve for the null geodesics analytically. It might be possible to learn more about the fate of infalling objects using these geodesics, since objects released sufficiently far away from the bottom of the throat will be approximately following a null geodesics by the time they reach the bottom. It may also be possible to construct Green's functions for these massless wave equations and study wave scattering in these geometries, investigations of this type have already been conducted for the $(1,0,n)$ family in 6 dimensions \cite{Heidmann:2019zws,Bena:2018bbd,Bena:2019azk}.

The ability to transform solutions so that the fluxes $\Theta^{(I)}$ are independent of $v$ is useful in its own right. Microstate geometries in general use a phenomena known as dissolving charges in fluxes to avoid having singular sources. To exploit this phenomena the fluxes need to thread non-trivial cycles in the geometry. In 5 dimensions the only non-trivial geometry is that of the GH base and so one can bring the full arsenal of tools developed for hyper-K\"{a}hler manifolds to study the $\Theta^{(I)}$, and by association through the BPS equations the rest of the data $(Z_{I},Z_{3},\Theta^{(3)},F,\omega)$. Using such transformations we showed how prepotentials can be constructed for the fluxes in 6 dimensions and explicitly constructed them for all $(k,m,n)$.

There are still open questions raised by the work of \cite{Tyukov:2018ypq} around whether it is possible to uncover a mathematical framework on hyper-K\"{a}hler manifolds that gives insight into BPS solutions? There it was shown how the $(\Theta^{I},\Theta^{(3)})$ fluxes are derived from prepotential functions and control moduli of the base which allow one to construct the $(Z_{I},Z_{3})$ analytically without solving any differential equations. The open questions are whether one can determine the moduli the $\Theta^{(I)}$ control? What are the new hyper-K\"{a}hler bases these moduli parametrize? As well as whether another principle can be found such that $\omega$ can be found without solving the final BPS equation? By demonstrating that the same tools can be used in 6 dimensions we have provided another setting in which these questions might be answered. Additionally the form of the prepotentials for general $(k,m,n)$ we constructed are richer than those known previously \cite{Tyukov:2018ypq}, perhaps they may shed light on some of these questions.  

It is hoped that the results presented here inform and motivate future study of the superstrata solutions, their rich structure promises to further the microstate geometry program and our understanding of black hole physics.

\section*{Acknowledgements}
I would like to thank Nicholas Warner, Pierre Heidmann and Felipe Rosso for valuable discussions and feedback on drafts or this work. The IPhT, CEA-Saclay provided accommodation and a stimulating environment during the genesis and completion of this project. This work was funded in part by the US Department of Energy under the grant DE-SC0011687 and by the ERC Grant 787320 - QBH Structure. 

\appendix
\section{Table 1: general structure of wave equations} \label{App: 1}

Here we present the general expressions of the 
\begin{align*}
G_{1}^{(6)}(r,\theta) =\frac{1}{2} \sum_{x_{1},x_{2}\in \mathcal{S}}x_{1}x_{2} G_{x_{1}x_{2}}(r,\theta)~,
\end{align*}
appearing in the wave equation (\ref{WEgen}) for superstrata obeying (\ref{rest1}) and (\ref{rest2}). The 5-dimensional result $G_{1}^{(5)}(r,\theta)=\left. G_{1}^{(6)}(r,\theta) \right|_{p=0}$ is given by omitting the last four rows.

\begin{table}[ht]
\begin{center}
\begin{tabular}{ |Sc|Sc| }
  \hline 
 $x_{1}x_{2}$ & $G_{x_{1}x_{2}}(r,\theta)$ \\ \hline \hline
$w^{2}$ &  $-\frac{\Lambda \Sigma}{4R^{2}r^{4}}\left[16 \Gamma \left(\frac{\omega_{\phi}-A_{\phi}\omega_{\psi}}{\sin 2\theta} \right)^{2} +r^{4}V\left(FP+V\omega_{\psi}^{2} \right) \right]$ \\ \hline
$w q_{1}$ &  $\frac{8\Gamma \Lambda \Sigma}{R r^{4}\sin^{2}2\theta} \left(\omega_{\phi} -A_{\phi}\omega_{\psi} \right)$ \\ \hline
$w q_{2}$ & $\frac{\Lambda\Sigma}{2Rr^{4}}\left( \frac{16A_{\phi}\Gamma(A_{\phi}\omega_{\psi}-\omega_{\phi})}{\sin^{2}2\theta}+r^{4}V^{2}\omega_{\psi} \right)$  \\ \hline
$q_{1}^{2}$ &  $-\frac{a^{4}\Gamma }{r^{4}} - \frac{4}{\sin^{2}2\theta}$ \\ \hline
$q_{2}^{2}$ & $-\frac{4\Gamma\Sigma\Lambda}{r^{4}}  \left(\frac{A_{\phi}}{\sin 2\theta}\right)^{2}- \frac{\Sigma \Lambda}{4} V^{2} $  \\ \hline
$q_{1}q_{2}$ &  $2\left( \frac{a^{4}\Gamma }{r^{4}} +\frac{1}{\sin^{2}\theta}+\frac{1}{\cos^{2}\theta} \right)A_{\phi}$ \\ \hline \hline
$p^{2} $ &  $- \frac{\Lambda \Sigma}{4R^{2}r^{4}} \left[ r^{4}V^{2}\beta_{\psi}^{2} +16 \Gamma \left( \frac{\beta_{\phi}-A_{\phi}\beta_{\psi}}{\sin 2\theta}\right)^{2}  \right]$  \\ \hline
$p w$ &  $ \frac{\Lambda\Sigma}{4R^{2}r^{4}} \left[ \frac{16 \Gamma \left(\beta_{\phi}-A_{\phi}\beta_{\psi} \right) \left( -\omega_{\phi}+A_{\phi}\omega_{\psi}\right)}{\sin^{2}2\theta}+r^{4}V \left( P-V\beta_{\psi}\omega_{\psi}\right)\right]$ \\ \hline
$pq_{1}$ & $\frac{8\Gamma \Lambda \Sigma}{R r^{4}\sin^{2}2\theta} \left(\beta_{\phi}-A_{\phi}\beta_{\psi} \right)$   \\ \hline
$p q_{2}$ &  $\frac{\Lambda \Sigma}{2r^{4}R}\left[ \frac{16 \Gamma A_{\phi}(A_{\phi}\beta_{\psi}-\beta_{\phi})}{\sin^{2}2\theta} +r^{4}V^{2}\beta_{\psi}\right]$ \\ \hline 
\end{tabular}
\caption{$G_{x_{1}x_{2}}(r,\theta)$ for superstrata with data satisfying (\ref{rest1}) and (\ref{rest2}).} \label{genericWave}

\end{center}
\end{table}

\newpage
\section{Table 2: $(1,0,n)$ and $(1,1,n)$ wave equations} \label{App: 2}
Here we present the  $G_{x_{1}x_{2}}(r,\theta)$ for the $(1,0,n)$ and $(1,1,n)$ original superstrata families (see section \ref{SubSec: wave equation structure} for definition of $G_{x_{1}x_{2}}(r,\theta)$), with
\begin{align*}
\Gamma=\frac{r^{2}}{a^{2}+r^{2}}~.
\end{align*}
If one makes the coordinate redefinition $\tilde{\theta}=\frac{\pi}{2}-\theta$, then it is clear the two families have identical 5-dimensional wave equations. This must be the case since they differ by a gauge transformation (\ref{GaugeTrans}) in 5 dimensions.

\begin{table}[ht] \label{10n table}
\begin{center}
\begin{tabular}{ |Sc|Sc| }
  \hline 
  $x_{1}x_{2}$ & $G_{x_{1}x_{2}}(r,\theta)$ \\ \hline \hline
$w^{2}$ &  $-\frac{\Gamma  \left(a^6+2 a^2 b^2 r^2 \left(\Gamma ^n-1\right)+b^4 r^2 \left(\Gamma ^n-1\right)\right)}{a^2 r^4}$ \\ \hline
$w q_{1}$ &  $\frac{2 \Gamma  \left(a^4+b^2 r^2 \left(\Gamma ^n-1\right)\right)}{r^4}$ \\ \hline
$w q_{2}$ & $\frac{2 \Gamma  \left((2 n+1) \left(a^4+r^2 \left(2 a^2+b^2\right)\right)-b^2 r^2 \Gamma ^n\right)}{r^4}$  \\ \hline
$q_{1}^{2}$ &  $-\frac{a^{4}\Gamma}{r^{4}}-\frac{4}{\sin^{2}2\theta}$ \\ \hline
$q_{2}^{2}$ & $-\frac{a^{4}(1+2n)^{2}\Gamma}{\mathbf{e}^{2}r^{4}}-\frac{4}{\mathbf{e}^{2}\sin^{2}2\theta} $  \\ \hline
$q_{1}q_{2}$ &  $-\frac{2 a^2 \Gamma  (2 n+1) \left(a^2+2 r^2\right)}{\mathbf{e}r^4} \pm\frac{8 \cos 2\theta}{\mathbf{e}\sin^{3}2\theta}$ \\ \hline \hline
$p^{2} $ &  $-\frac{a^{4}(\mathbf{a}-\mathbf{e}+2\mathbf{a}n)^{2}\Gamma}{4n^{2}r^{4}}-\frac{(\mathbf{a}-\mathbf{e})^{2}}{n^{2}\sin^{2}2\theta}$  \\ \hline
$p w$ &  $   \pm \frac{(\mathbf{a}-\mathbf{e}+2\mathbf{a}n)(a^{4}+(2a^{2}+b^{2})r^{2})\Gamma}{nr^{4}}\mp\frac{b^{2}(\mathbf{a}-\mathbf{e)}\Gamma^{1+n}}{nr^{2}}$ \\ \hline
$pq_{1}$ & $\mp\frac{a^{2}(\mathbf{a}-\mathbf{e}+2\mathbf{a}n)(\Gamma+1)}{nr^{2}} + \frac{4(\mathbf{a}-\mathbf{e})\cos 2\theta}{n \sin^{2}2\theta}$   \\ \hline
$p q_{2}$ &  $\mp\frac{a^{4}(1+2n)(\mathbf{a}-\mathbf{e}+2\mathbf{a}n)\Gamma}{\mathbf{e}nr^{4}}\mp \frac{4(\mathbf{a}-\mathbf{e})}{\mathbf{e}n\sin^{2}2\theta}$ \\ \hline 
\end{tabular}
\caption{$G_{x_{1}x_{2}}(r,\theta)$ for the $(1,0,n)$ family (upper sign) and $(1,1,n)$ family (lower sign).}
\end{center}
\end{table}

\newpage
\begin{landscape}
\section{Table 3: $(2,1,n)$ wave equations} \label{App 3}
Here we present the  $(G^{(0)}_{x_{1}x_{2}},G^{(b)}_{x_{1}x_{2}},G^{(c)}_{x_{1}x_{2}})$ for the $(2,1,n)$ original and supercharged superstrata families.
\begin{table}[ht] \label{21n table}
\begin{center}
\begin{tabular}{ |Sc|Sc|Sc|Sc| }
  \hline 
 $x_{1}x_{2}$ & $G^{(0)}_{x_{1}x_{2}}(r,\theta)$ & $G^{(b)}_{x_{1}x_{2}}(r,\theta)$& $G^{(c)}_{x_{1}x_{2}}(r,\theta)$   \\ \hline \hline
$w^{2}$ & $-\frac{a^{4}\Gamma}{r^{4}}$ & $h_{b}(r)$& $h_{c}(r)$\\ \hline
$w q_{1}$ & $\frac{2 a^{4}\Gamma}{r^{4}}$ &  $-\frac{2b^{2}\Gamma}{r^{2}}+\frac{b^{2}\left(a^{2}(3+2n)+2r^{2} \right)\Gamma^{2+n}}{r^{4}}$& $\frac{2 c^2 \Gamma  \left(-a^4 (n (n+2)-2)+6 a^2 r^2+4 r^4\right)}{a^4 n (n+2) r^2}-\frac{(\Gamma +1) \Gamma ^{n+1} \left(a^2 c (n+2)+2 c r^2\right)^2}{a^4 n (n+2) r^2}$\\ \hline
$w q_{2}$ & $\frac{4a^{2}(1+n)(1+\Gamma)}{\mathbf{e}r^{2}}$ & $\frac{4 b^2 \Gamma  (n+1)}{\mathbf{e} r^2}+\frac{2 a^2 b^2 (n+1) \Gamma ^{n+2}}{\mathbf{e} r^4}$& $\frac{4 c^2 \Gamma  (n+1) \left(a^2 (n (n+2)+2)+2 r^2\right)}{a^2 \mathbf{e} n (n+2) r^2}-\frac{2 (n+1) \Gamma ^{n+2} \left(a^2 c (n+2)+2 c r^2\right)^2}{a^2 \mathbf{e} n (n+2) r^4}$ \\ \hline
$q_{1}^{2}$ & $-\frac{a^{4}\Gamma}{r^{4}} - \frac{4}{\sin^{2}2\theta}$ & $0$& $0$ \\ \hline
$q_{2}^{2}$ & $-\frac{4a^{4}(1+n)^{2}\Gamma}{\mathbf{e}^{2}r^{4}}$ & $0$& $0$\\ \hline
$q_{1}q_{2}$ & $-\frac{4a^{2}(1+n)(1+\Gamma)}{\mathbf{e}r^{2}}$ & $0$& $0$\\ \hline \hline
$p^{2} $ & $-\frac{a^4 \Gamma  (\mathbf{e}-2 \mathbf{a} (n+1))^2}{4 (n+1)^2 r^4}-\frac{\mathbf{e}^2}{(n+1)^2\sin^{2}2\theta}$ & $0$& $0$ \\ \hline
$p w$ &  $\frac{a^2 (\Gamma +1) (2 \mathbf{a} (n+1)-\mathbf{e})}{(n+1) r^2}$ & $\frac{b^2 \Gamma  (2 \mathbf{a} (n+1)-\mathbf{e}) \left(a^2 \Gamma ^{n+1}+2 r^2\right)}{2 (n+1) r^4}+\frac{a^2 b^2 \mathbf{e}  \Gamma ^{n+2} \cos 2 \theta }{r^4}$& $\frac{c^2 \Gamma  (2 \mathbf{a} (n+1)-\mathbf{e}) \left(\Gamma ^{n+1} \left(-\left(a^2 (n+2)+2 r^2\right)^2\right)+2 a^2 (n (n+2)+2) r^2+4 r^4\right)}{2 a^2 n (n+1) (n+2) r^4}$ \\ \hline
$pq_{1}$ & $\frac{a^2 (\Gamma +1) (\mathbf{e}-2 \mathbf{a} (n+1))}{(n+1) r^2}-\frac{4 \mathbf{e} \cos 2\theta}{(n+1)\sin^{2}2\theta}$ & $0$& $0$ \\ \hline
$p q_{2}$ &  $\frac{2 a^4 \Gamma  (\mathbf{e}-2 \mathbf{a} (n+1))}{\mathbf{e} r^4}$ & $0$& $0$ \\ \hline 
\end{tabular}
\caption{$G_{x_{1}x_{2}}(r,\theta)$ for the $(2,1,n)$ original and supercharged families.}
\end{center}
\end{table}
\begin{align*}
h_{b}(r)&=  \frac{2 a^2 b^2+b^4}{a^2 r^2} \Gamma -\frac{b^2 \left(2 a^4 (n+2)+a^2 \left(b^2 (n+1)+4 r^2\right)+2 b^2 r^2\right)}{2 a^2 r^4} \Gamma^{2+n} -\frac{b^4}{4 r^4} \Gamma^{3+2n}~,\\
h_{c}(r)&= \frac{2 a^4 c^4 (n (n+2)-2) r^2-8 a^2 c^4 r^4+a^6 c^2 n (n+2) \left(2 a^2 n (n+2)+c^2 (n (n+2)+2)\right)-4 c^4 r^6}{a^8 n^2 (n+2)^2 r^2} \Gamma  -\frac{\left(a^2 c (n+2)+2 c r^2\right)^4}{4 a^8 n^2 (n+2)^2 r^4} \Gamma^{3+2n} \\
& \qquad \qquad \qquad  \qquad  \qquad -\frac{c^2 \left(a^2 (n+2)+2 r^2\right) \left(2 a^4 c^2 \left(n^2-4\right) r^2-4 a^2 c^2 (n+4) r^4+a^6 n (n+2)^2 \left(2 a^2 n+c^2 (n+1)\right)-8 c^2 r^6\right)}{2 a^8 n^2 (n+2)^2 r^4} \Gamma^{2+n} ~.
\end{align*}
\end{landscape}

\begin{adjustwidth}{-1mm}{-1mm} % to adjust the L and R margins 
\bibliographystyle{utphys}      
\bibliography{microstates}       % calls file "microstates.bib"

\end{adjustwidth}

%%%%%%%%%%%%%%%%%%%
%%%%%%%%%%%%%%%%%%%

\end{document}